\documentclass[12pt]{article}
\usepackage{graphicx,amsmath,amssymb,amsfonts}
\usepackage{citesort,epsfig}
\usepackage{url}

\begin{document}

\begin{titlepage}
\begin{centering}
\vfill

{\large \bf New avenue to the Parton Distribution Functions: Self-Organizing Maps}

\vspace{0.5cm}
 H. Honkanen$^{\rm a,b,}$\footnote{heli@iastate.edu}, 
 S. Liuti$^{\rm a,}$\footnote{sl4y@virginia.edu} \\
 
\vspace{0.3cm}
{\em $^{\rm a}$Department of Physics, University of Virginia,
P.O. Box 400714, Charlottesville, VA 22904-4714, USA}\\
\vspace{0.3cm}
{\em $^{\rm b}$Department of Physics and Astronomy, Iowa State University,\\
Ames, IA 50011, USA}\\
\vspace{0.3cm}

\vspace{0.3cm}
 J. Carnahan$^{\rm c}$,
%\footnote{joe.carnahan@gmail.com},
 Y. Loitiere$^{\rm c}$,
 P.~R. Reynolds$^{\rm c}$
\\
 
\vspace{0.3cm}
{\em $^{\rm c}$Department of Computer Science,
School of Engineering, \\
University of Virginia, P.O. Box 400740
Charlottesville, VA 22904-4740, USA}\\

\vspace{0.5cm}\centerline{\bf Abstract}
\end{centering}

%%%%%% ABSTRACT:
Neural network algorithms have been recently applied to construct Parton
Distribution Function (PDF) parametrizations which provide an alternative
to standard global fitting procedures. We propose a technique based on an
interactive neural network algorithm using Self-Organizing Maps (SOMs).
SOMs are a class of clustering algorithms based on 
competitive learning
among spatially-ordered neurons. Our SOMs are trained on 
selections of stochastically generated PDF samples. The selection
criterion for every optimization iteration is based on the features
of the clustered PDFs. 
Our main goal is to provide a fitting procedure that, at variance with
the standard neural network approaches, allows for an increased control of
the systematic bias by enabling  user interaction in the various stages of
the process.

\vfill
\end{titlepage}

\section{Introduction}

Modelling experimental data always introduces 
bias, in the form of either a theoretical or systematical bias. The former
is introduced by researchers with the precise structure of the 
model they use, which invariably constrains the form of the solutions.
The latter form of bias is introduced by algorithms, such as 
optimization algorithms, 
which may favour some results in ways which are not justified by their 
objective
functions, but rather depend on the internal operation of the algorithm.

In this paper we concentrate on high energy hadronic interactions, which are 
believed to be described by Quantum Chromodynamics (QCD). 
Because of the properties of factorization and asymptotic freedom of the 
theory, 
the cross sections for a wide number of hadronic reactions can be computed
using perturbation theory, as convolutions of perturbatively calculable 
hard scattering coefficients, with non perturbative Parton Distribution 
Functions (PDFs) that 
parametrize the large distance hadronic structure.  
The extraction of the PDFs from experiment is inherently 
affected by a bias, which ultimately dictates the accuracy
with which the  theoretical predictions can be compared 
to the high precision 
measurements of experimental observables.
In particular, the form of bias introduced by PDFs  will necessarily 
impact the upcoming searches of physics beyond the Standard 
Model at the Large Hadron Collider (LHC). This situation has in fact 
motivated an impressive body of work, and continuous, ongoing 
efforts to both estimate and control PDFs uncertainties.

%%%%%%%%%%%%%%%%%%%%%%%%%%%%%%%%%%%%%%%%%%%%%%%%%%%%
%  GLOBAL ANALYSIS
%%%%%%%%%%%%%%%%%%%%%%%%%%%%%%%%%%%%%%%%%%%%%%%%%%%%
Currently, the established method to obtain the 
PDFs is the global analysis, a fitting procedure,
where initial scale 
$ Q_0\sim 1 {\mbox{GeV}}\le Q_{\rm dat}^{\rm min}$
ansatze, as a function of the momentum fraction $x$,
%, such as $ f_{i/h}(x,{Q_0})={a_0}x^{a_1}(1-x)^{a_2}P(x;a_3,...)$ 
for each parton flavour $i$ in hadron $h$ are evolved to higher scales
according to the perturbative QCD renormalization group equations. 
All the available observables {\it e.g.} 
the proton structure function,
$F_2^p(x,Q^2)$, are composed of the candidate PDFs and comparison with 
the data is made with the help of some statistical estimator such as the global
$\chi^{2}$,
\begin{equation}
\chi^{2}=\sum_{\mathrm{expt.}}\sum_{i,j=1}^{N_{e}} 
\left({\rm Data}_{i}-{\rm Theor}_{i}\right)
V_{ij}^{-1}\left({\rm  Data}_{j}-{\rm Theor}_{j}\right),
\end{equation}
where the error matrix $V_{ij}$ consists of the statistical
and uncorrelated systematic errors, as well as of the
correlated systematic errors when available.
The parameters in the ansatze are then adjusted and the whole process repeated 
until a global minimum has been found.

The modern PDF collaborations (CTEQ \cite{Nadolsky:2008zw} and references 
within , MRST
\cite{Martin:2001es,Martin:2004ir,Martin:2007bv}, 
Alekhin \cite{Alekhin:2002fv,Alekhin:2006zm}, Zeus \cite{Chekanov:2005nn}
and H1 \cite{Adloff:2003uh})
also provide error
estimates for the PDF sets.
They all rely on some kind of variant of
 the Hessian method (see e.g. \cite{Pumplin:2001ct}
for details), which is based on a Taylor
expansion of the global $\chi^2$ around it's minimum.
When only the leading 
terms are kept, the displacement of $\chi^2$ can be written in terms of Hessian
matrix $H_{ij}$, 
which consists of second derivatives of $\chi^2$ with respect to the
parameter displacements, evaluated at the minimum.
The error estimate for the parameters themselves, or for any quantity
 that depends 
on those parameters, can then be obtained in terms of the inverse of the 
Hessian matrix,
\begin{equation}
(\Delta X)^2={\Delta\chi^2}\sum_{i,j}\frac{\partial X}{\partial y_i}
\left( H^{-1}\right)_{ij}\frac{\partial X}{\partial y_j}.
\end{equation}
For details of PDF uncertainty studies  see e.g. 
Refs.~\cite{Martin:2003sk,Martin:2002aw}.

The global analysis combined with Hessian error estimation
 is a powerful method, allowing for both extrapolation 
outside the kinematical range of the data and extension to multivariable cases,
 such as nuclear PDFs (nPDFs) 
\cite{Eskola:2008ca,Eskola:2007my,Hirai:2004wq,Hirai:2007sx}.
In principle, when more data become available,
 the method could also be applied to Generalized Parton 
Distributions (GPDs), for which only model-dependent 
\cite{Vanderhaeghen:1999xj}
or semi model-dependent \cite{Ahmad:2006gn,Ahmad:2007vw}
solutions presently exist.

However, there are uncertainties related to the method itself, that are 
difficult to quantify, but may turn out to have a large effect.
Choosing global $\chi^2$ as a  statistical estimator may not be
adequate since the minimum of the global fit may not correspond to a
minima of the individual data sets, and as a result the
definition of  $\Delta \chi^2$ may be ambiguous.  Estimates for the current 
major global analyses are that 
$\Delta \chi^2 = 50 - 100$ is needed to obtain a $\sim 90\%$ confidence 
interval  \cite{Nadolsky:2008zw,Martin:2001es}. In principle 
this problem could be avoided by using the Lagrange multiplier method
(see e.g.\cite{Stump:2001gu}),
which does not assume quadratic behaviour for the errors around the minimum,
instead of the Hessian method, but this is computationally more expensive
solution.
Introducing a functional form
at the initial scale necessarily introduces a parametrization dependence bias
and theoretical assumptions behind the fits, such as  
 $s$, $\bar s$, $c$ quark content, 
details of the scale evolution (e.g. higher order perturbative corrections, 
large/small $x$ 
resummation), higher twists
etc. as well as the data selection and treatment, e.g. kinematical cuts,
all reflect into the final result of the analysis. Also, there may
be systematical bias introduced by the optimization algorithm. 
The differences between the current global PDF sets tend to be larger than
the estimated uncertainties \cite{Pumplin:2005yfa},
and these differences again translate to the predictions for the LHC 
observables, such as Higgs  \cite{Djouadi:2003jg} or $W^\pm$ and $Z$ production
cross sections  \cite{Nadolsky:2008zw}.

%%%%%%%%%%%%%%%%%%%%%%%%%%%%%%%%%%%%%%%%%%%%%%%%%%%%
%  NEURAL NETWORK APPROACH
%%%%%%%%%%%%%%%%%%%%%%%%%%%%%%%%%%%%%%%%%%%%%%%%%%%%

A new, fresh approach to the PDF fitting has recently been proposed by
NNPDF collaboration \cite{Ubiali:2008uk,Ball:2008by}
who have replaced a typical functional form ansatz with
a more complex  standard neural network (NN) solution and the Hessian method
with Monte Carlo sampling of the data (see 
the references within  \cite{Ball:2008by} for the nonsinglet PDF fit and
the details of the Monte Carlo sampling).

%%%%%%%%%%%%%%%%%%%%%%%%%%%%%%%%%%%%%%%%%%%%%%%%%%%%%%%%%%%%%%%%%%%%%%%
\begin{figure}[h]
\begin{center}
\vspace{-0.2cm}
\epsfysize=3cm\epsffile{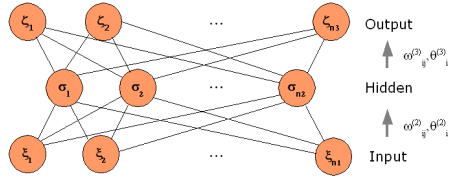} \hspace{-1.0cm}
\vspace{-0.2cm}
\caption[a]{\protect \small Schematic diagram of a feed-forward neural network,
from \cite{DelDebbio:2007ee}. }
\label{nn.eps}
\end{center}
\end{figure}
\vspace{-0.0cm}
%%%%%%%%%%%%%%%%%%%%%%%%%%%%%%%%%%%%%%%%%%%%%%%%%%%%%%%%%%%%%%%%%%%%%%%

Neural network can be described as a computing solution that consists of 
interconnected processing 
elements, neurons, that work together 
to produce an output function.
In a typical feed-forward NN (see Fig.~\ref{nn.eps})
the output is given by the neurons in the last layer, as a non-linear
function of the 
output of all neurons in the  previous layer, which in turn is a function of 
the output of all neurons in the previous layer, and so on, starting from the 
first layer, which receives the input. For a NN with $L$ layers and $n_l$ 
neurons in each layer, the total number of the parameters is
$\sum_{l=1}^{L-1}\left( n_l\,n_{l+1}\,+n_{l+1}\right)$.

In the beginning of the NNPDF fitting procedure
a Monte Carlo sample of replicas of the experimental data is generated 
by jittering the central values of the data withing their errorbars 
using 
univariate Gaussian (or some other distribution if desired)
random numbers for each independent error source.
The number of the replicas is made so large that the Monte Carlo set of
replicas models faithfully the probability distribution of the original data.
For each replica a Genetic Algorithm (GA) fit is performed by first
setting the NN parameters for each parton flavour to be fitted randomly, then
making clones of the set of parameters, and mutating
each of them randomly (multiple mutations). After scale evolution the 
comparison with the data is performed for all the clones, and
the best clones are selected for
a source of new clones,
and the process repeated until the minimum for the $\chi^2$ has been found.
Overfitting of the data is prevented by using only part of the data in the
minimizing procedure, and using the other part to monitor the behaviour
of the $\chi^2$.
When fitting PDFs one thus ends up with $N_{\rm rep}$ PDF sets, each initial 
scale parton distribution parametrized by a different NN.
The quality of the global fit is then given by the 
$\chi^2$ computed from the
averages over the sample of trained neural networks.
The mean value of  the parton distribution at the starting scale
for a given value of $x$ is found by averaging over the replicas,
and the uncertainty on this value is the variance of the values 
given by the replicas.

The NNPDF method circumvents the problem of choosing a suitable 
$\Delta \chi^2$, and it relies on GA which 
works on a population of solutions for each MC replica, thus having
a lowered possibility of getting trapped in local minima.
NN parametrizations are also highly complex, with large number of parameters,
and thus unbiased compared to the ansatze used in global fits.
The estimated uncertainties for NNPDF fits are larger than those of global
fits, possibly indicating that the global fit uncertainties may have been
underestimated. It should, however, be pointed out that the MC sampling of
the data is not not tied to use of NNs, and it thus remains undetermined 
whether the large uncertainties would persist if the MC sampling was used with 
a fixed functional form.
The complexity  of NN results may also induce problems, especially when used in
a purely automated fitting procedure. Since the effect of modifying individual
NN parameters is unknown, the result may exhibit strange or unwanted behaviour
in the extrapolation region, or in between the data points if the data is 
sparse. In such a case, and in a case of incompatible data, the overfitting
method is also unsafe to use.
Implementation of information not given directly by the data, such
as nonperturbative models, lattice calculations or  knowledge from prior work
in general, is also difficult in this approach. 

A possible method of estimating the PDF uncertainties could also be provided
by Bayesian statistical analysis, as preliminarily studied in 
\cite{Giele:2001mr,Cowan:2006fi} and explained in \cite{d'agostini},
in which the errors for the PDF parameters, or for an observable constructed 
from the PDFs, are first encapsulated in prior probabilities for an enlarged
set of model parameters, and posterior distributions are obtained using 
computational tools such as Markov Chain Monte Carlo. Similar to NNPDF 
approach, this method allows for an inclusion of 
non-Gaussian systematic errors for the data.

In this introductory paper we propose a new method which relies on the use of 
Self-Organizing Maps (SOMs), a subtype of neural network.
The idea of our method is to create means for introducing
``Researcher Insight'' instead of ``Theoretical bias''. In other
words, we want to give up fully automated fitting procedure and eventually 
develop an interactive
fitting program which would allow us to ``take the best of both worlds'',
to combine the best features of both the standard functional form approach and 
the neural network approach. In this first step, we solely concentrate on 
single
variable functions, free proton PDFs, but discuss the extension of the model 
to multivariable cases.
In Section~\ref{SOM method} we describe the general features of the SOMs,
in Sections~\ref{MIXPDF} and \ref{ENVPDF} we present two PDF fitting
algorithms relying on the use of SOMs and finally in Section~\ref{future} 
we envision the possibilities the SOM method has to offer.

%%%%%%%%%%%%%%%%%%%%%%%%%%%%%%%%%%%%%%%%%%%%%%%%%%%%
%  SOM DESCRIPTION
%%%%%%%%%%%%%%%%%%%%%%%%%%%%%%%%%%%%%%%%%%%%%%%%%%%%

\section{Self-Organizing Maps}
\label{SOM method}

The SOM is a visualization algorithm which
attempts to represent all the available observations with optimal accuracy 
using a restricted set of models.
The SOM was developed by T.~Kohonen in the early 1980's (\cite{Teuvo},
see also \cite{somtoolbox})
to model biological brain functions, but has since then developed into a 
powerful computational approach on it's own right.
Many fields of science, such as statistics, signal processing, control theory, 
financial analyses, experimental physics, chemistry and medicine,
have adopted the SOM as a standard analytical tool.
SOMs have been applied to texture discrimination, 
classification/pattern recognition, motion detection,
genetic activity mapping, drug discovery, cloud classification, and
speech recognition, among others. Also,
a new application area is organization of very large document collections. 
However, applications in particle physics have been scarce so far, and mostly 
directed to improving the algorithms for background event rejection
\cite{Lange:1997qh,Becks:1999fa,Lange:1999mb}.

SOM consists of nodes, map cells, which are all assigned spatial 
coordinates, and  the topology of the map is determined by a chosen distance 
metric $M_{\mathrm{map}}$. Each cell $i$ contains  a  map vector $V_i$,
that is isomorphic to the 
data samples used for training of the neural network. 
In the following we will concentrate on a 2-dimensional 
rectangular lattice for simplicity. 
A natural choice for the topology is then 
$L_1(x,y)=\sum_{i=1}^2\vert x_i-y_i\vert$, 
which also has been proved \cite{Aggawal} to be an ideal choice for 
high-dimensional data, such as PDFs in our case.

\begin{figure}[h]
\begin{center}
\vspace{-0.2cm}
\epsfysize=4.5cm\epsffile{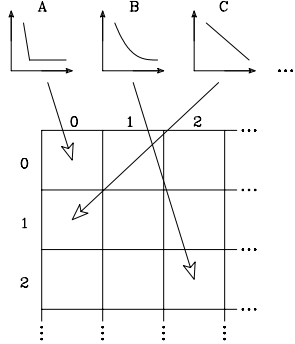} \hspace{-1.0cm}
\vspace{-0.2cm}
\caption[a]{\protect \small A 2D grid SOM which cells get randomly associated
with the type of data samples we would like to study, such as nonsinglet 
PDFs or observables. At this stage each cell gets associated with only one
 curve, the map vector.}
\label{train1.ps}
\end{center}
\end{figure}

The implementation
 of SOMs proceeds in three stages: 1) initialization of the SOM,
2) training of the SOM  and 3) associating the data samples with a
trained map, i.e. clustering.
During the initialization the map vectors are chosen such that
each cell is set to contain an arbitrarily selected sample of either the 
actual data to be clustered, or anything isomorphic to them 
(see Fig.~\ref{train1.ps} for an 
example). The actual training data samples, which may be e.g. subset or the
whole set of the actual data,
 are then associated with map vectors
by minimizing a  similarity metric $M_{\mathrm{data}}$. We choose 
$M_{\mathrm{data}}=L_1$. 
The map
vector each data sample becomes matched against, is then the most similar one 
to the data sample among  all the other map vectors.
It may happen that some map vectors do no not have any samples 
associated with them, and some may actually have many.

%%%%%%%%%%%%%%%%%%%%%%%%%%%%%%%%%%%%%%%%%%%%%%%%%%%%%%%%%%%%%%%%%%%%%%%
\vspace{-0.0cm}
%%%%%%%%%%%%%%%%%%%%%%%%%%%%%%%%%%%%%%%%%%%%%%%%%%%%%%%%%%%%%%%%%%%%%%%

%%%%%%%%%%%%%%%%%%%%%%%%%%%%%%%%%%%%%%%%%%%%%%%%%%%%%%%%%%%%%%%%%%%%%%%
\begin{figure}[h]
\begin{center}
\vspace{-0.2cm}
\epsfysize=7cm\epsffile{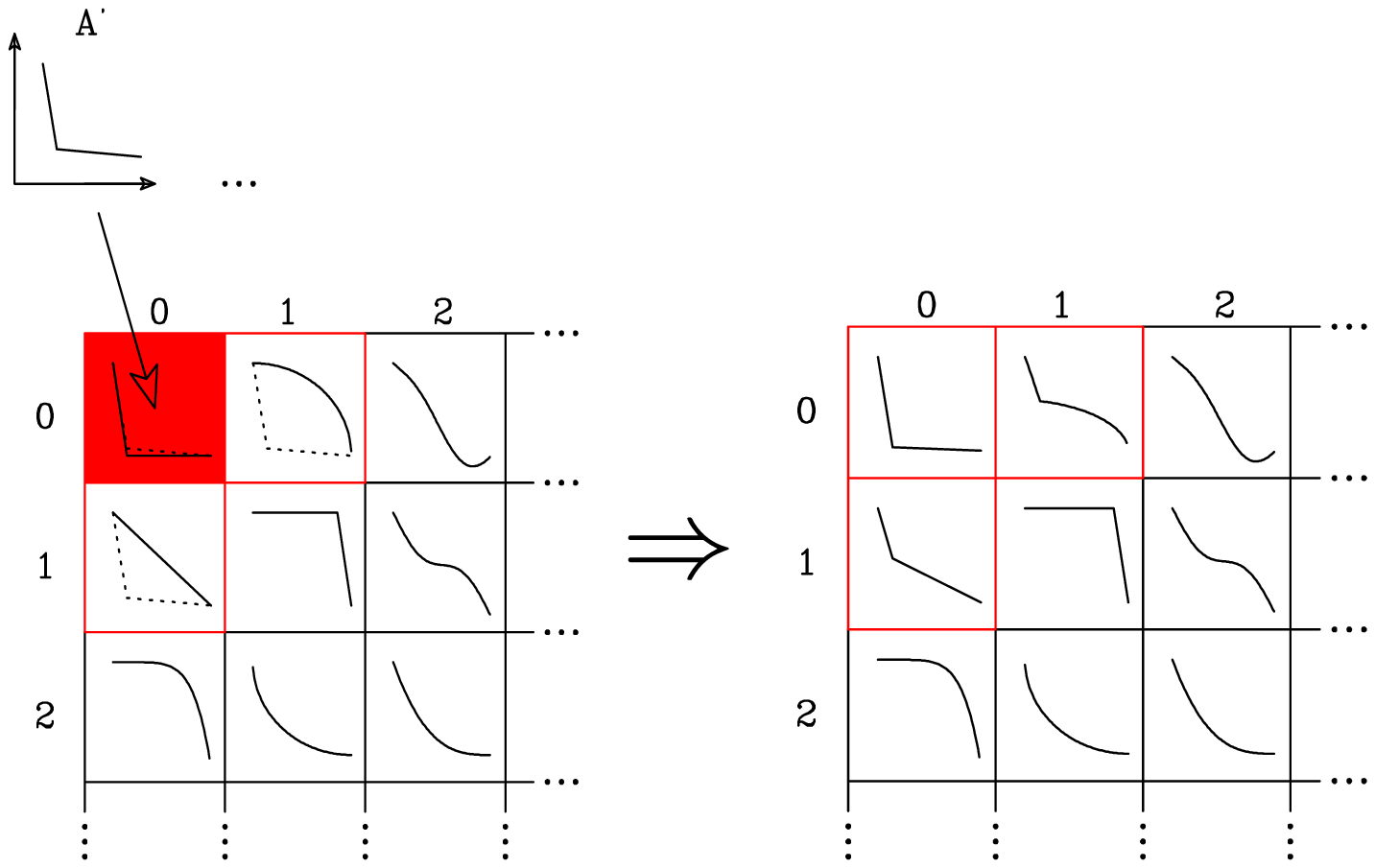} \hspace{-1.0cm}
\vspace{-0.5cm}
\caption[a]{\protect \small (Colour online)
 Each data sample $S_i$ is associated with the 
one map vector $V_i$ it is most similar to. As a reward for the match, the 
winning map vector, as well as its neighbouring map vectors,
get averaged with the data associated with the winning cell.}
\label{train2.eps}
\end{center}\end{figure}
\vspace{-0.0cm}
%%%%%%%%%%%%%%%%%%%%%%%%%%%%%%%%%%%%%%%%%%%%%%%%%%%%%%%%%%%%%%%%%%%%%%%

During the training  the map vectors 
are updated by averaging them with the 
data  samples that fell into the cells within a given decreasing 
neighbourhood, see 
Fig.~\ref{train2.eps}. This type of training which is based on rewarding the
winning node to become more like data, is called {\it competitive learning}.  
The initial value
of a map vector $V_i$ at SOM cell $i$ 
then changes during the course of
training as
\begin{equation}
V_{i}(t+1) = V_i(t) \, \left( 1-w(t) \, N_{j,i}(t)\right) 
+ S_j(t) \, w(t) \, N_{j,i}(t)
\label{somtrain}
\end{equation}
where now $V_{i}(t+1)$ is the contents of the SOM cell $i$ after the
data sample $S_j$ has been presented on the map.
The neighbourhood, the radius, within
which the map vectors are updated is given by the function $N_{j,i}(t)$, 
centered on the winner cell $j$. 
Thus even the map
vectors in those cells that didn't find a matching data sample are adjusted, 
rewarded, to become more like data.
Typically
$N_{j,i}(t) = e^{-M_{map}(j,i)^2 / r(t) }$,
where 
$r(t)$ is a monotonously decreasing radius sequence. In the beginning of the 
training the neighbourhood may contain the whole map 
and in the end it just consists of the cell itself.
Moreover, the updating 
is also controlled by 
$w(t)$, which is a monotonously decreasing weight sequence in the range 
$[0,1]$.

As the training proceeds the neighbourhood 
function eventually causes the data samples to be placed on a certain region 
of the map,
where the neighbouring map vectors are becoming increasingly similar to each
other, and the weight sequence $w(t)$ furthermore finetunes their position.

In the end on a properly trained SOM, cells  that are topologically close to 
each other will have map vectors which are similar to each other.
In the final phase the actual data is matched against the 
map vectors of the trained map, and thus get distributed on the map according 
to the feature that was used as $M_{\mathrm{data}}$. Clusters with similar
data now emerge as a result of {\it unsupervised learning}.

\begin{figure}[h]
\begin{center}
\vspace{-0.2cm}
\epsfysize=4.5cm\epsffile{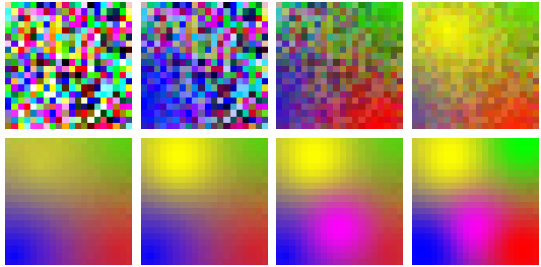} \hspace{-1.0cm}
\vspace{-0.6cm}
\caption[a]{\protect \small (Colour online)
SOM containing RGB colour triplets getting trained. Adapted from 
Ref.~\cite{RGB-sites}}
\label{somcolors.eps}
\end{center}
\end{figure}

For example, a map containing RGB colour triplets would initially have
colours randomly scattered around it, but during the course of
training it would evolve into patches
of colour which smoothly blend with each other, see Fig.~\ref{somcolors.eps}.
This local similarity property is the feature that makes SOM suitable for 
visualization purposes, 
thus facilitating user interaction with the data. Since each map vector now
represent a class of similar objects, the SOM
is an ideal tool to
visualize high-dimensional data, by projecting it onto a low-dimensional map
clustered according to some desired similar feature. 
%%%%%%%%%%%%%%%%%%%%%%%%%%%%%%%%%%%%%%%%%%%%%%%%%%%%%%%%%%%%%%%%%%%%%%%
\vspace{-0.0cm}
%%%%%%%%%%%%%%%%%%%%%%%%%%%%%%%%%%%%%%%%%%%%%%%%%%%%%%%%%%%%%%%%%%%%%%%

SOMs, however, also have disadvantages. Each SOM, though identical in size and
shape and containing same type of data, is different. The clusters may also 
split in such a way that similar type of data can be found in several 
different places on the map. We are not aware of any 
mathematical or computational means  of 
detecting if and when the map is fully trained, and whether there occurs
splitting or not, 
other than actually computing the similarities between
the neighbouring cells and studying them.

In this work we use the so-called batch-version of the training, in which
all the training data samples are matched against the map vectors before the
training begins. The map vectors 
 are then averaged with all the training  samples within the neighbourhood 
radius simultaneously.
 The procedure is repeated  $N_{\mathrm{step}}$ (free parameter to 
choose) times such that in
every training step the {\it same} set of training data samples is associated
with the evolving map and in Eq.(\ref{somtrain}) $t$ now counts
training steps.
 When the map is trained, the actual data is finally
matched against the map vectors. In our study our training data are always
going to be the whole data we want to cluster, and the last training step
is thus the clustering stage.
The benefit of the batch training compared to the incremental training,
described earlier, is that the training is independent of the order in which
the training samples are introduced on the map.

%%%%%%%%%%%%%%%%%%%%%%%%%%%%%%%%%%%%%%%%%%%%%%%%%%%%
%  MIXPDF
%%%%%%%%%%%%%%%%%%%%%%%%%%%%%%%%%%%%%%%%%%%%%%%%%%%%

\section{MIXPDF algorithm}
\label{MIXPDF}

In this approach our aim is to both i) to be able to study the 
properties of the PDFs in a model independent way and yet ii) to be able to 
implement knowledge from the prior works on PDFs, and ultimately iii)
to be able to
guide the fitting procedure interactively with the help of SOM properties.

At this stage it is important to distinguish between the experimental data
and the training data of the SOM. When we are referring to measured data
used in the PDF fitting, such as $F_2$ data, we always call it experimental
data. The SOM training data in this study is going to be a collection of
candidate PDF sets, produced by us, or something composed of them.
A PDF set in the following will always mean a set of 8 curves,
one for each independent parton flavour $f=(g, u_v, d_v, 
\bar u, \bar d, s={\bar s}, c={\bar c}$ and $ b={\bar b}$ in this simplified
introductory study), 
that are properly normalized such that
\begin{equation}
\sum_f\int_0^1 dx x f_{f/p}(x,Q^2)=1,\label{momsumrule}
\end{equation}
and  conserve baryon number and charge
\begin{equation}
\int_0^1 dx f_{u_v/p}(x,Q^2)=2,\,
\int_0^1 dx f_{d_v/p}(x,Q^2)=1.
\label{baryonsumrule}
\end{equation}

In order to proceed we have to decide how to create our candidate PDF sets,
decide the details of the SOMs, details of the actual fitting algorithm, 
experimental data selection and details of the scale evolution.

In this introductory paper our aim is not to provide a finalised SOMPDF
set, but rather to explore the possibilities and restrictions of the
method we are proposing. Therefore we refrain from using ``all the possible
experimental data'' as used in global analyses, but concentrate on
%``care free''
DIS structure function data from H1 \cite{Adloff:2000qk}, 
BCDMS \cite{Benvenuti:1989rh,Benvenuti:1989fm} and Zeus \cite{Chekanov:2001qu},
which we use without additional kinematical cuts or normalization factors
(except rejecting the data points below our initial scale).
The parameters for the DGLAP scale evolution were chosen to be those of
 CTEQ6 (CTEQ6L1 for lowest order (LO)) \cite{Pumplin:2002vw:cteq6}, 
the initial scale being 
$Q_0=1.3$ GeV.
In next-to-leading order
(NLO) case the evolution code was taken from \cite{QCDNUM} 
(QCDNUM17 beta release).

We will start now with a simple pedagogical example, which we call MIXPDF 
algorithm, where we use some of the 
existing PDF sets as material for new candidate PDFs.
At first, we will choose
CTEQ6 \cite{Pumplin:2002vw:cteq6}, 
CTEQ5 \cite{Lai:1999wy:cteq5},
MRST02 \cite{Martin:2001es,Martin:2002dr}, Alekhin \cite{Alekhin:2002fv} and 
GRV98 \cite{Gluck:1998xa} sets and construct new PDF sets from them such
that at the initial scale each parton flavour in the range $x=[10^{-5},1]$
 is randomly selected from one of these five sets (we set 
the heavy flavours
to be zero below their mass thresholds). The  sumrules on this new
set are then imposed such that the original normalization of $u_v$ and $d_v$
are preserved, but the rest of the flavours are scaled 
together so that Eq.(\ref{momsumrule}) is fulfilled. In this study we accept
the $<$few\% normalization error which results from the fact that our 
x-range is not $x=[0,1]$. From now on we call these type of PDF sets 
{\it database} PDFs.
We randomly initialize a small $5\times 5$ map with these
candidate database PDFs,
such that each map vector $V_i$
consists of the PDF set itself, and
of the observables $F_2^p(x,Q_0^2)$  derived from it.
Next we train the map with  $N_{\mathrm{step}}=5$ batch-training steps
with training data that consists of 100 database PDFs plus 5 original
``mother'' PDF sets, which we will call {\it init} PDFs from now on. We choose
the similarity criterion to be the 
%added 
similarity of observables 
$F_2^p(x,Q^2)$
% and $F_2^d(x,Q^2)$ 
with
$M_{\mathrm{data}}=L_1$. The similarity is tested at a number of 
$x$-values (equidistant in logarithmic scale up to $x\sim 0.2$, and equidistant
in linear scale above that) 
both at the initial scale and at all the evolved scales where experimental data
exist.
On every training, after the matching, all 
the observables (PDFs) of the map vectors get averaged with the observables 
(PDFs, flavor by flavor) matched within the neighbourhood according to 
Eq.~(\ref{somtrain}). 
The resulting new averaged map
vector PDFs are rescaled again (such that  $u_v$ and $d_v$ are scaled first)
to obey the sumrules. From now on we will call
these type of PDF sets {\it map} PDFs. 
The map  PDFs  are evolved and the 
observables at every experimental data scale are computed
and compared for similarity with  the observables from the 
training PDFs.

\begin{figure}[h]
\begin{center}
\vspace{-0.2cm}
\hspace*{-2.0cm}
\epsfysize=9cm\epsffile{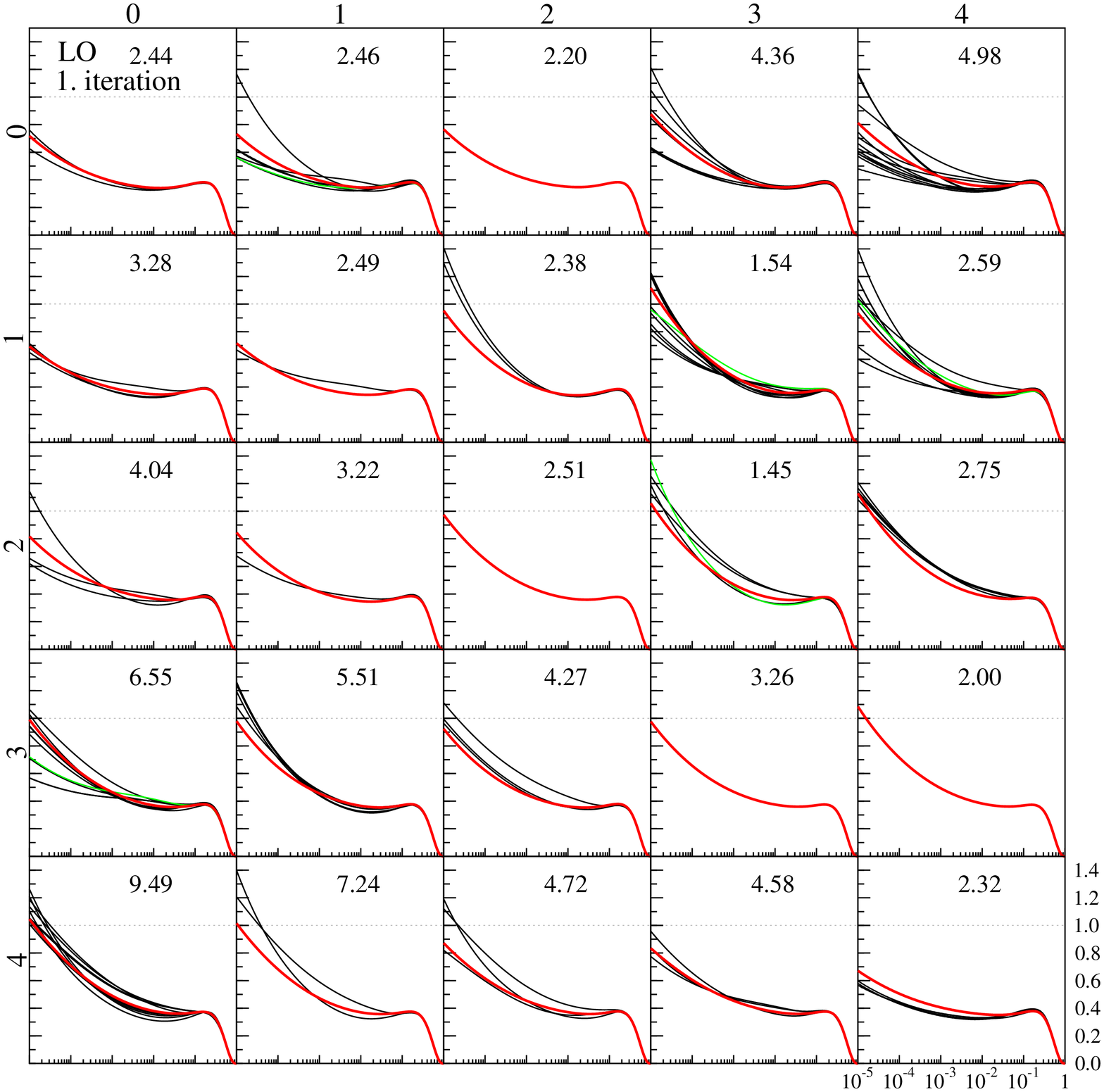} 
\vspace{-.5cm}
\caption[a]{\protect \small (Colour online)
Trained 1. iteration LO map.
Red curves are map $F_2^p$'s, 
green curves init $F_2^p$'s and  black curves rest of the 
database $F_2^p$'s. The number in each cell is the $\chi^2/N$ of the  
map PDF. }
\label{mappdf_shot_1.eps}
\end{center}
\end{figure}

After the training we have a map with 25 map PDFs and the same 105 PDF sets we 
used to train the map. 
The resulting LO SOM is shown in 
Fig.~\ref{mappdf_shot_1.eps},
with just
$F_2^p(x,Q_0^2)$'s of each cell shown for clarity. The red curves in the figure
are the map F2's constructed from the map PDFs, black curves are the database
F2's and green curves are the init F2's constructed from the init PDFs
(CTEQ6, CTEQ5, MRST02, Alekhin and GRV98 parametrizations).
It is obviously difficult to judge visually
just by looking at the curves whether the map is good and fully trained.
One hint about possible ordering may be provided by the fact that the
shape of the $F_2^p$ curve must correlate with the $\chi^2/N$ against the
experimental data. The 
distribution of the $\chi^2/N$ values (no correlated systematic
 errors are taken 
into account for simplicity)
of the map PDFs, shown in each cell , does indeed seem somewhat organized. 
%%%%%%%%%%%%%%%%%%%%%%%%%%%%%%%%%%%%%%%%%%%%%%%%%%%%%%%%%%%%%%%%%%%%%%%
\vspace{-0.0cm}
%%%%%%%%%%%%%%%%%%%%%%%%%%%%%%%%%%%%%%%%%%%%%%%%%%%%%%%%%%%%%%%%%%%%%%%%

Table~\ref{mixmothertab} lists the  $\chi^2/N$ values the original
init PDFs obtain within the MIXPDF framework described above. Comparison of
these values with the values in 
Fig.~\ref{mappdf_shot_1.eps} reveals that some of the map PDFs,
as also some of the database PDFs, have gained a
$\chi^2/N$ comparable to or better than that of the init PDFs.

%\renewcommand{\thefootnote}{\fnsymbol{footnote}}

%%%%%%%%%%%%%%%%%%%%%%%%%%%%%%%%%%%%%%%%%%%%%%%%%%%%%%%%%%%%%%%%%%%%%%%
\begin{table}[h]
\center
\begin{tabular}{|c|c|c|}
\hline
PDF\footnotemark & LO $\chi^2/N$ & NLO $\chi^2/N$\\ \hline
Alekhin &  3.34 & 29.1 \\ \hline
CTEQ6 & 1.67 & 2.02 \\ \hline
CTEQ5 & 3.25 & 6.48  \\ \hline
CTEQ4 & 2.23 & 2.41 \\ \hline
MRST02  & 2.24 & 1.89 \\ \hline
GRV98 & 8.47 & 9.58 \\ \hline
\end{tabular}
\caption{$\chi^2/N$ for different MIXPDF input PDF sets 
 against all the datasets 
used (H1, ZEUS, BCDMS, N=709).}
\label{mixmothertab}
\end{table}

%%%%%%%%%%%%%%%%%%%%%%%%%%%%%%%%%%%%%%%%%%%%%%%%%%%%%%%%%%%%%%%%%%%%%%%

\begin{figure}[h]
\begin{center}
\vspace{-0.2cm}
\hspace*{-1.5cm}
\epsfysize=8cm\epsffile{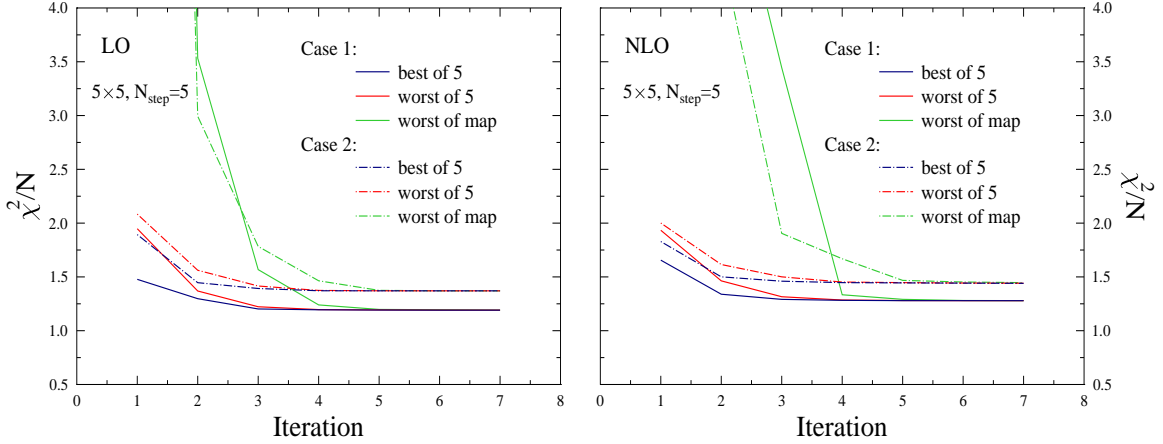} 
\vspace{-1.5cm}
\caption[a]{\protect \small (Colour online)
$\chi^2/N$ of the MIXPDF runs as a function
of the iteration. }
\label{chi2_mappdf.eps}
\end{center}
\end{figure}

\begin{figure}[h]
\begin{center}
\vspace{-0.2cm}
\hspace*{-1.0cm}
\epsfysize=9cm\epsffile{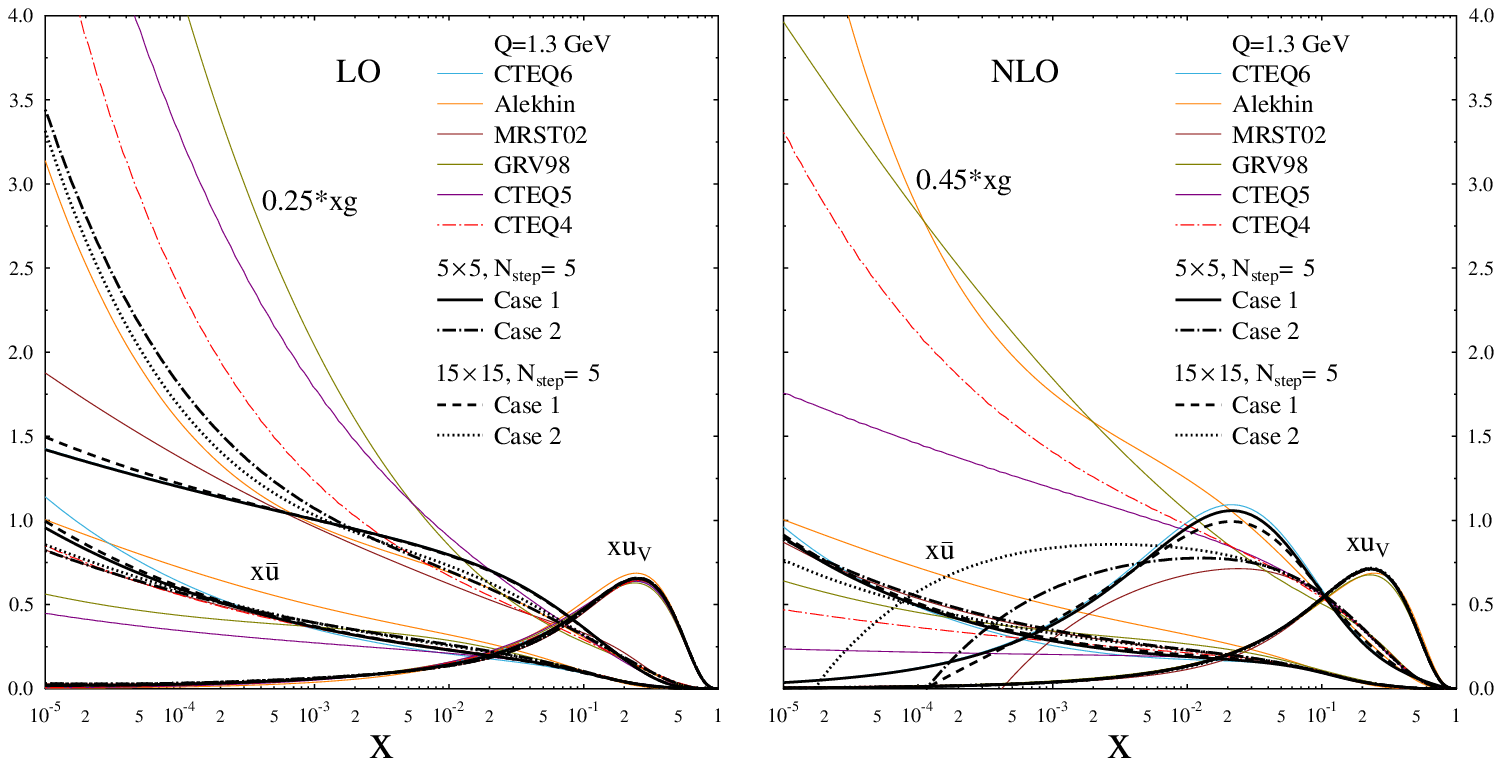} \hspace{-0.0cm}
\vspace{-1.5cm}
\caption[a]{\protect \small (Colour online)
MIXPDF results together with the input PDF sets. }
\label{mappdf_jakaumat.eps}
\end{center}
\end{figure}

Inspired by the progress, we start a second
fitting {\it iteration} by selecting the 5 best PDF sets from the 25+5+100
PDF sets of the first iteration as our new init PDFs (which are now properly
normalized after the first iteration) to 
generate database PDFs for a whole new 
SOM. Since the best PDF 
candidate from the first  iteration is matched on this new map as an unmodified
init PDF, it is guaranteed that the  $\chi^2/N$ as a function of the
iteration either decreases or remains the same.  
We keep repeating the 
iterations until the $\chi^2/N$ saturates. 
Fig.~\ref{chi2_mappdf.eps} (Case 1) shows the  $\chi^2/N$ as a 
function of iterations for the best PDF on the trained map, for the worst 
PDF on the map and for the worst of the 5 PDFs selected for the subsequent 
iteration as an init PDF.
The final $\chi^2/N$ of these runs are listed in Table~\ref{mixpdfvarietytab}
(first row) as
Case 1  and Fig.~\ref{mappdf_jakaumat.eps} shows these results 
(black solid line), 
together with
original starting sets at the initial scale (note the different scaling
for gluons in LO and in NLO figures). 
For 10 repeated LO runs we obtain  $\bar{\chi^2}/N$=1.208 and 
$\sigma$=0.029.

\footnotetext{These are the  $\chi^2/N$
for the initial scale
PDF sets taken from the quoted parametrizations and
evolved with CTEQ6 DGLAP settings, the heavy flavours were set
to be zero below their mass thresholds, no kinematical cuts
or normalization factors for
the experimental data were imposed, and no correlated systematic 
 errors of the data
were used to compute the  $\chi^2/N$. 
We do not claim these 
values to describe the quality of the quoted PDF sets.}
%\renewcommand{\thefootnote}{\arabic{footnote}}

%%%%%%%%%%%%%%%%%%%%%%%%%%%%%%%%%%%%%%%%%%%%%%%%%%%%%%%%%%%%%%%%%%%%%%
\vspace{-0.0cm}
%%%%%%%%%%%%%%%%%%%%%%%%%%%%%%%%%%%%%%%%%%%%%%%%%%%%%%%%%%%%%%%%%%%%%%%
%%%%%%%%%%%%%%%%%%%%%%%%%%%%%%%%%%%%%%%%%%%%%%%%%%%%%%%%%%%%%%%%%%%%%%%
\begin{table}[h]
\center
\begin{tabular}{|c|c|c|c|c|c|c|}
\hline
SOM & $N_{\rm step}$ & \# init & \# database & Case
& LO $\chi^2/N$ & NLO $\chi^2/N$\\ \hline
5x5 & 5  & 5 & 100 & 1 & 1.19  & 1.28 \\  \hline
5x5 & 5  & 5 & 100 & 2 & 1.37  &  1.44 \\ \hline
5x5 & 5  & 10 & 100 & 1 & 1.16 & 1.25 \\ \hline
5x5 & 5  & 10 & 100 & 2 & 1.49 & 1.43 \\ \hline
5x5 & 5  & 15 & 100 & 1 & 1.16 & 1.45 \\ \hline
5x5 & 5  & 20 & 100 & 1 & 1.17 & - \\ \hline
5x5 & 10  & 10 & 100 & 1 & 1.16 & 1.30 \\ \hline
5x5 & 40  & 10 & 100 & 1 & 1.20 & - \\ \hline
15x15 & 5  & 5 & 900 & 1 & 1.22 & - \\ \hline
15x15 & 5  & 5 & 900 & 2 & 1.31 & - \\ \hline
15x15 & 5  & 30 & 900 & 1 & 1.16 & 1.25 \\ \hline
15x15 & 5  & 30 & 900 & 2 & 1.25 & l.53 \\ \hline
\end{tabular}
\caption{$\chi^2/N$ against all the datasets 
used (H1, ZEUS, BCDMS) for some selected MIXPDF runs. }
\label{mixpdfvarietytab}
\end{table}

%%%%%%%%%%%%%%%%%%%%%%%%%%%%%%%%%%%%%%%%%%%%%%%%%%%%%%%%%%%%%%%%%%%%%%%
%%%%%%%%%%%%%%%%%%%%%%%%%%%%%%%%%%%%%%%%%%%%%%%%%%%%%%%%%%%%%%%%%%%%%%%
\vspace{-0.0cm}
%%%%%%%%%%%%%%%%%%%%%%%%%%%%%%%%%%%%%%%%%%%%%%%%%%%%%%%%%%%%%%%%%%%%%%%

\begin{figure}[h]
\begin{center}
\vspace{-0.2cm}
\hspace*{-2.0cm}
\epsfysize=9cm\epsffile{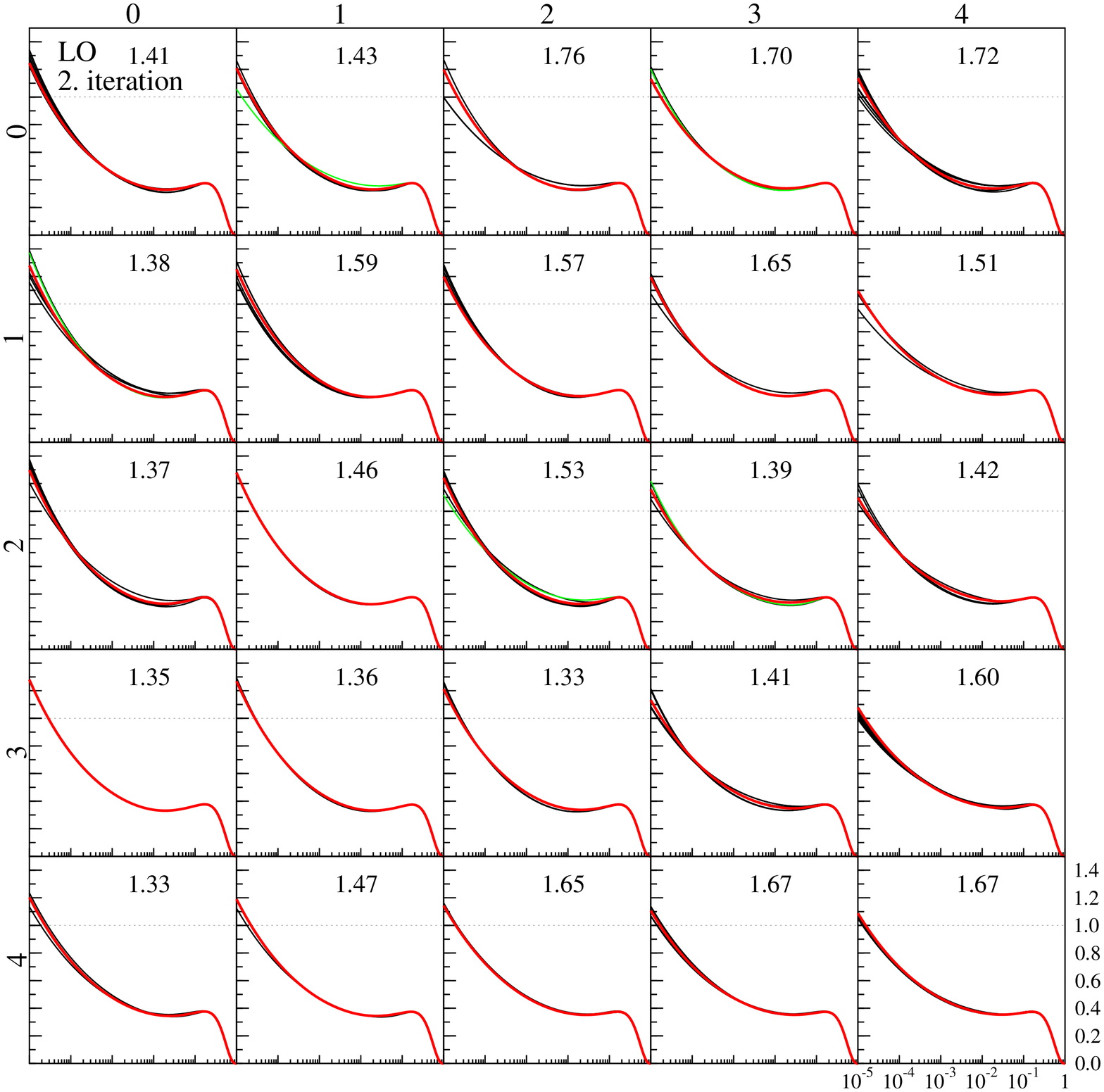} 
\vspace{-.5cm}
\caption[a]{\protect \small (Colour online)
Trained 2. iteration LO map, curves and numbers
as in Fig.\ref{mappdf_shot_1.eps}. }
\label{mappdf_shot_2.eps}
\end{center}
\end{figure}

\begin{figure}[h]
\begin{center}
\vspace{-0.2cm}
\hspace*{-2.0cm}
\epsfysize=9cm\epsffile{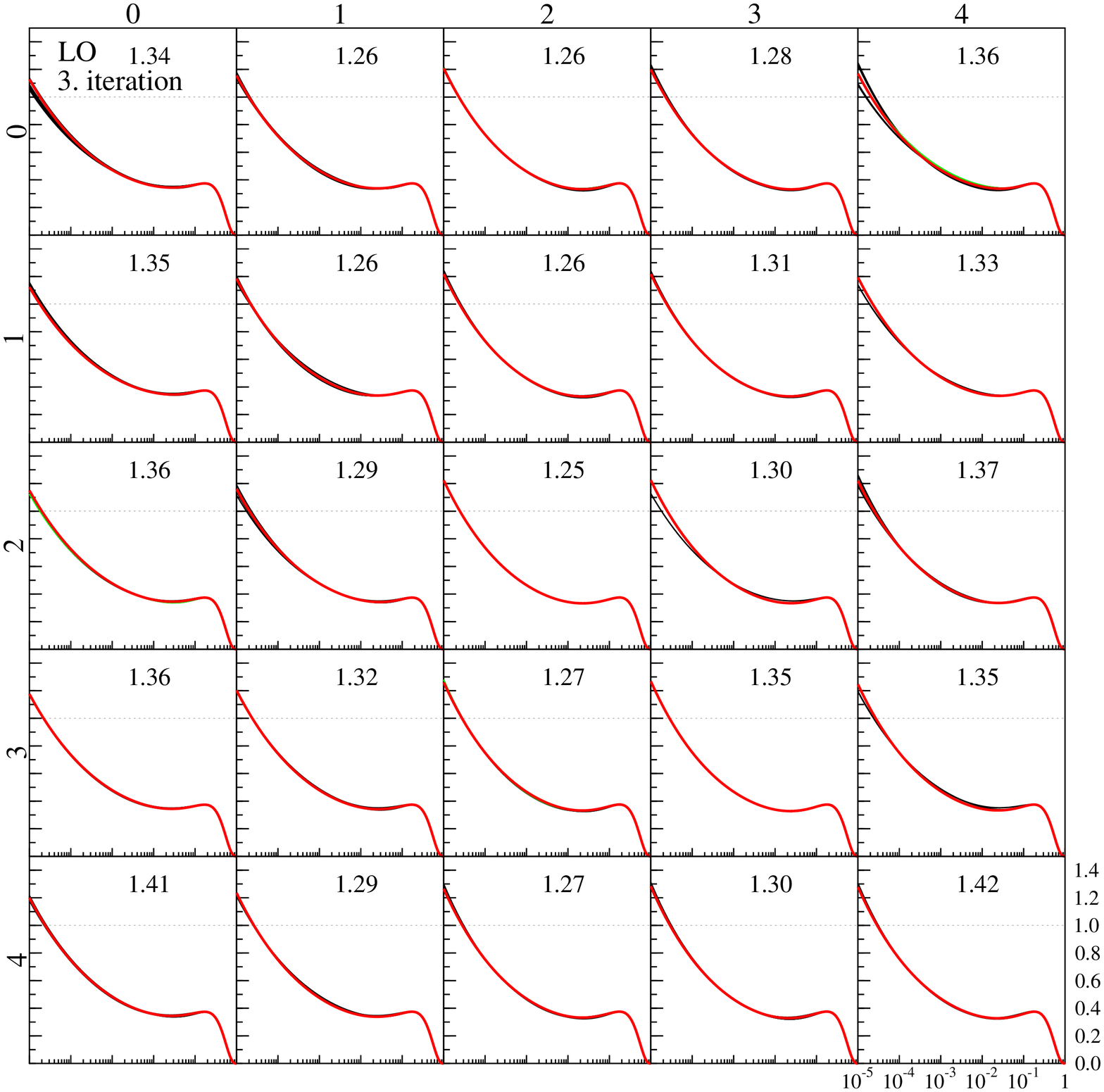} 
\vspace{-.5cm}
\caption[a]{\protect \small (Colour online)
Trained 3. iteration LO map, curves and numbers
as in Fig.\ref{mappdf_shot_1.eps}. }
\label{mappdf_shot_3.eps}
\end{center}
\end{figure}

Let us now analyze in more detail how the optimization proceeds.
Figs.~\ref{mappdf_shot_2.eps},\ref{mappdf_shot_3.eps} show the LO maps
also for the 2. and 3. iterations. 
On the first iteration the init PDFs, the shapes of which were taken from 
existing parametrizations,
fall in the cells (0,1) (CTEQ5), (1,3) (Alekhin), (1,4) (MRST02), (2,3)
(CTEQ6) and (3,0) (GRV98), so the modern sets, Alekhin, MRST02 and CTEQ6,
group close to each other, i.e. the shapes of the observables they produce
are very similar, as expected. The best 5 PDFs selected as the 2. iteration
init PDFs 
also come from this neighbourhood, 3 of them from the cell (1,3) and the other
2 from the cell (2,3). Two of these selected sets are map PDFs, two are 
database
PDFs and also the original init PDF CTEQ6 survived for the 2. iteration.

At the end of the 2. iteration the init PDFs, which originated from
the neighbouring cells, are scattered in the cells (0,1), (0.3), (1,0) (CTEQ6),
(2,2) and (2,3) and the best 5 PDFs produced during this iteration are in the
cells (4,1) (database PDF), (4,0) (map PDF), (3,2) (map PDF), 
(2,2) (database PDF) and (3,0) (map PDF).

After the 3. iteration the above best 5 PDFs of 2. iteration are in the cells
(2,0), (3,1), (0.2), (0.4) and (3,2) and the new best 5 PDFs produced are 
all map PDFs with 4 of them in neighbouring cell pairs.

Map PDFs provide  complicated linear combinations
of the database PDFs and obviously
play an important role in the algorithm.
The size of the map dictates how much the neighbouring map vectors differ
from each other. Since the PDFs in the same cell are not required to be 
similar, only the observables constructed from them are, a cell or a 
neighbourhood may in principle contain a spread of PDFs with a spread of  
$\chi^2/N$'s. 
However, since our selection criteria for the init PDFs was
based on the best $\chi^2/N$ only, it is inevitable that the 
observables on 
the map
become increasingly similar as the iterations go by, and the 
$\chi^2/N$ flattens very fast as can be seen from 
Fig.~\ref{chi2_mappdf.eps}.
As a consequence
we quickly lose the variety in the shapes of the PDFs 
as the iterations 
proceed, and on the final iteration all the PDFs on the map end up being
very similar.

%%%%%%%%%%%%%%%%%%%%%%%%%%%%%%%%%%%%%%%%%%%%%%%%%%%%%%%%%%%%%%%%%%%%%%%
\vspace{-0.0cm}
%%%%%%%%%%%%%%%%%%%%%%%%%%%%%%%%%%%%%%%%%%%%%%%%%%%%%%%%%%%%%%%%%%%%%%%%
%%%%%%%%%%%%%%%%%%%%%%%%%%%%%%%%%%%%%%%%%%%%%%%%%%%%%%%%%%%%%%%%%%%%%%%
\vspace{-0.0cm}
%%%%%%%%%%%%%%%%%%%%%%%%%%%%%%%%%%%%%%%%%%%%%%%%%%%%%%%%%%%%%%%%%%%%%%%%

The MIXPDF algorithm obviously has several other weaknesses too.
 Among them are
how the result would change if we started with another
first iteration PDF selection, and
what are the effects of changing
the size of the map, number of the database PDFs and init PDF 
sets and size of 
$N_{\mathrm{step}}$? In general, how much the final result depends on the 
choices that we make during the fitting process? 

Let us now study some of these questions a bit. 
Since we have chosen our evolution 
settings to be those of CTEQ6's, it necessarily becomes a favoured set 
(although
we don't impose any kinematical cuts on the experimental data).
Therefore we performed another LO and NLO runs, with CTEQ6 now replaced 
with CTEQ4 
\cite{Lai:1996mg:cteq4}. The results of these runs are reported
in  Table~\ref{mixpdfvarietytab} (2. row) and in 
Fig.~\ref{chi2_mappdf.eps}  ($\chi^2/N$)
and Fig.~\ref{mappdf_jakaumat.eps} (the PDFs) as Case 2 . The Case 2
clearly produces worse results.
Without an input from CTEQ6 we automatically lose all the
low gluons at small-$x$ -type of results in NLO, for example.

\begin{figure}[h]
\begin{center}
\vspace{-0.2cm}
\hspace*{-1.5cm}
\epsfysize=8cm\epsffile{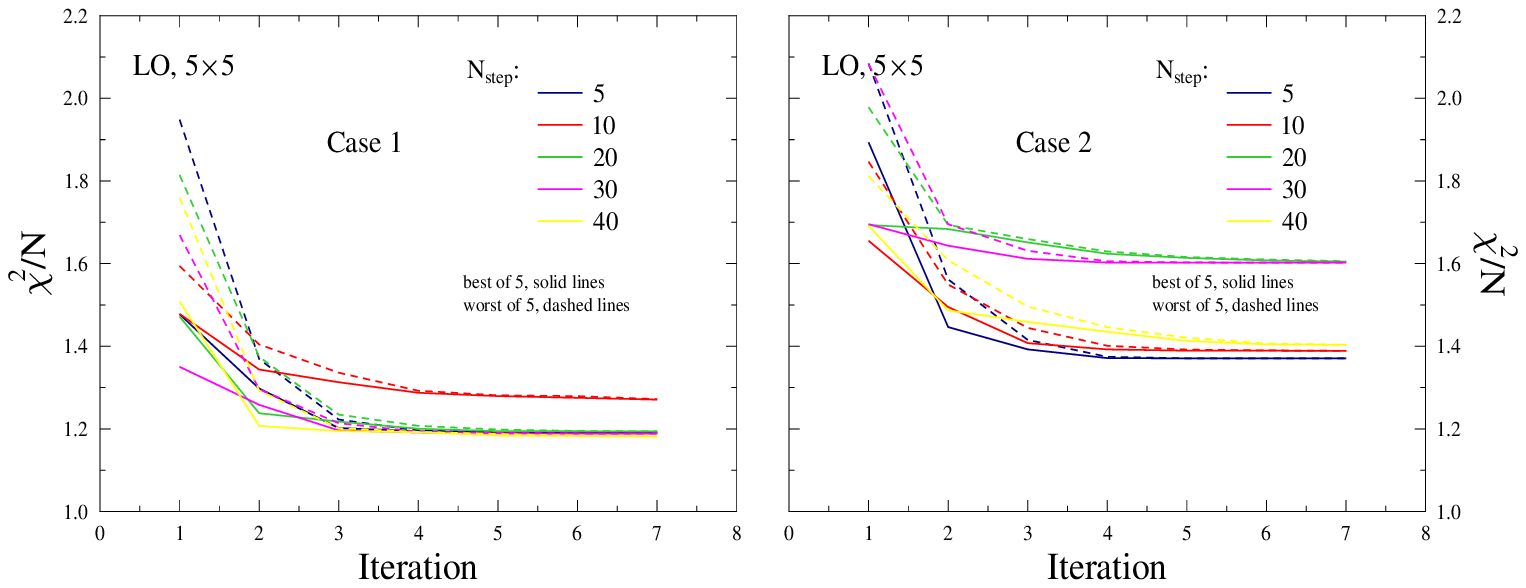}
\vspace{-1.5cm}
\caption[a]{\protect \small (Colour online)
LO $\chi^2/N$ for $5\times 
5$ SOM runs with different
$N_{\mathrm{step}}$. }
\label{chi2_cases.eps}
\end{center}
\end{figure}

Fig.~\ref{chi2_cases.eps} addresses the issue of choosing different 
$N_{\mathrm{step}}$, for LO Case 1 and Case 2. 
The solid (dotted) lines
show the best (worst) init PDF selected from each iteration for several
$N_{\mathrm{step}}$ selections. In Case 2 the best results are obtained with
small number of training steps, whereas Case 1 does not seem to benefit from
a longer SOM training.  
Keeping the stochastical nature of the process in our minds, we may
speculate that the seemingly opposite behaviour for
the Case 1 and Case 2 results from the fact that it is more probable to
produce a good set of database PDFs in Case 1 than in Case 2.
If the database
is not so good to begin with, the longer training  contaminates all the map
PDFs with the low quality part of the database PDFs.
%%%%%%%%%%%%%%%%%%%%%%%%%%%%%%%%%%%%%%%%%%%%%%%%%%%%%%%%%%%%%%%%%%%%%%%
\vspace{-0.0cm}
%%%%%%%%%%%%%%%%%%%%%%%%%%%%%%%%%%%%%%%%%%%%%%%%%%%%%%%%%%%%%%%%%%%%%%%

Table~\ref{mixpdfvarietytab} also showcases best
results from a variety of MIXPDF runs 
where we have
tried different combinations of SOM features. It is interesting to notice that
increasing the size of the map and database does not necessarily lead to a 
better performance. 
Instead the number of the init PDFs on later iterations (always 5 for the
first iteration)
seem to be a key
factor, it has to be sufficiently large to preserve the variety of database 
the PDFs 
but small enough to consist of PDFs with good quality.
 In our limited space of database candidates 
($\sim 6^5=7776$ possibilities) the optimal 
set of variables for the MIXPDF run should not be not impossible to map down.

The method we used to generate the sample data is very simple indeed.
The number of all the possible candidate database PDFs is not very large to 
begin with, so the quality of the final results strongly depends on the 
quality of the input for the SOM. 
Since the map PDFs are obtained
by averaging with the training samples,   and  the non-valence
flavours are scaled by a common factor when imposing the sumrules,
the map PDFs tend to lie in between the init PDFs.
Therefore map PDFs with extreme shapes are never produced, and thus never
explored by the algorithm.

A method which relies on sampling existing parametrizations on a SOM is 
inconvenient also because it is not readily applicable to multivariable cases.
For the case of the PDFs it is sufficient to have a value for each flavour
for a discrete set of $x$-values, but for a multivariable cases, such as
nPDFs, or GPDs the task of keeping track of the grid of values for each flavour
in each $x$ for several different 
values of the additional variables ({\it e.g.} $A$ and $Z$ for nuclei, 
and the skewness, $\xi$ and the squared 4-momentum transfer,
$t$ for GPDs) is computationally expensive.
In principle, SOM can keep track of the interrelations of the map vectors, and
knowing the parametrization for the init PDFs, it would be possible to
construct the parametrization for the map PDFs.
That would, however, lead to very complicated parametrizations, and 
different nPDF, GPD etc. parametrizations  are presently not even either 
measured or defined at the same initial scale.

Despite of its problems, on a more basic level, MIXPDF does have the
the desirable feature  that it allows us
to use SOM as a part of the PDF optimization algorithm 
in such a way that we cluster our candidate PDFs on the map, and select
those PDFs which {\it i)} minimize a chosen fitness function, e.g. $\chi^2$, 
when
compared  against experimental data, and {\it ii)} have some desired feature
which can be visualized on the map or used as a clustering criterion.
Therefore, in the following Section, we keep building on this example.

\section{ENVPDF algorithm}
\label{ENVPDF}

Most of the problems with the MIXPDF algorithm  originate from the need
to be able to generate the database PDFs in an unbiased way as possible, and 
at the same time to have a variety of PDF candidates available at every stage 
of the fitting procedure.
Yet, one needs to have control over the features of the database PDFs that are 
created.

To accomplish this, we choose, at variance with the ``conventional'' PDFs
sets or NNPDFs, to give up the functional form of PDFs 
and rather to rely on purely stochastical methods in 
generating the initial and training samples of the PDFs. Our choice is
a GA-type analysis, in which our parameters are the values of PDFs at the 
initial scale 
for each flavour at each value of $x$ where the experimental data exist.
To obtain control over the shape of the PDFs  we use some of the
existing distributions 
to establish an initial range, or {\em envelope}, within which we sample the 
database PDF values. 

Again, we use the Case 1 and 2 PDF sets 
(CTEQ6, CTEQ5, CTEQ4, MRST02, Alekhin and GRV98) 
as an initialization guideline. We construct our initial PDF generator first
to, 
for each flavour separately, 
select randomly  either the range  $[0.5,1]$, 
$[1.0,1.5]$ or $[0.75,1.25]$ times any of the Case 1 (or 2) PDF set.
Compared to MIXPDF algorithm we are thus adding more freedom to the scaling
of the database PDFs.
Next the initial generators  generate values for  
each $x_{\rm data}$\footnote{
To ensure a reasonable large-$x$ behaviour for the PDFs, we also generate
with the same method
values for them in a few $x$-points outside the range of the experimental
data. We also require the PDFs, the gluons especially, to be positive for 
simplicity.}
 using uniform, instead of Gaussian, 
distribution around the existing parametrizations, thus reducing
 direct bias from them.
Gaussian smoothing is applied to the 
resulting set of points, and  the flavours 
combined to form a PDF set such that the curve is linearly 
interpolated from the discrete set of generated points.

The candidate PDF sets are
then scaled to obey the sumrules as in MIXPDF algorithm. 
In order to obtain a reasonable selection of PDFs to start with, we reject 
candidates which have $\chi^2/N>10$ (computed as in MIXPDF algorithm).
To further avoid direct bias from the Case 1 and 2 PDFs, we don't include 
the init PDFs
into the training set for the first iteration as we did in MIXPDF case.
For a $N\times N$ SOM we choose the size of the database to be $4N^2$.
 
During the later iterations we proceed as follows:
At the end of each iteration we pick from the trained $N\times N$ SOM $2N$ 
best PDFs as the init PDFs.
These init PDFs are introduced into the training set alongside with the 
database PDFs, which are now constructed using each of the init PDFs 
{\it in turn} as a center for a Gaussian random number generator, which 
assigns for {\it all} the flavours for each $x$ a value around that {\it same}
 init PDF
such that $1-\sigma$ of the generator is given by the spread of the best PDFs 
in the topologically nearest neighbouring cells. 
The object of these generators is thus to refine a good candidate PDF found 
in the previous iteration by jittering it's values within a  range
determined by the shape of other good candidate PDFs from the previous 
iteration.
The generated PDFs are then 
smoothed and
scaled to obey the sumrules. Sets with  $\chi^2/N>10$ are always rejected.
We learnt from the MIXPDF algorithm that it is important to preserve
the variety of the PDF shapes on the map, so
we also keep $N_{\rm orig}$ copies of the first iteration 
generators in our generator mix.

Table~\ref{envpdftab} lists results from a variety of such runs. The results
do not seem to be very sensitive to the number of SOM training steps,
$N_{\rm step}$, but are highly sensitive to the number of first iteration 
generators used in subsequent iterations. Although the generators can 
now in principle produce an infinite number of different PDFs, the algorithm 
would not be able
to radically change the shape of the database PDFs without introducing
a random element on the map. Setting $N_{\rm orig}> 0$ provides,
through map PDFs, that element, and keeps the algorithm from
getting fixed to a local minimum. 
The ENVPDF algorithm is now more independent
from the initial selection of the PDF sets, Case 1 or 2, than MIXPDF, 
since no values of {\it e.g.} the CTEQ6 set in the original 
generator  are ever introduced on the map directly.

%%%%%%%%%%%%%%%%%%%%%%%%%%%%%%%%%%%%%%%%%%%%%%%%%%%%%%%%%%%%%%%%%%%%%%%
\begin{table}[h]
\center
\begin{tabular}{|c|c|c|c|c|c|c|}
\hline
SOM & $N_{\rm step}$ & $N_{\rm orig}$ & Case
& LO $\chi^2/N$ & NLO $\chi^2/N$\\ \hline
5x5 & 5  & 2 & 1 & 1.04 & 1.08  \\ \hline
5x5 & 10  & 2 & 1 & 1.10 & - \\ \hline
5x5 & 20  & 2 & 1 & 1.10 & - \\ \hline
5x5 & 30  & 2 & 1 & 1.10 & - \\ \hline
5x5 & 40  & 2 & 1 & 1.08 & - \\ \hline
5x5 & 5  & 0 & 1 & 1.41 & - \\ \hline
5x5 & 20  & 0 & 1 & 1.26 & - \\ \hline
5x5 & 5  & 2 & 2 & 1.14 & 1.25 \\ \hline
5x5 & 10  & 2 & 2 & 1.12 & - \\ \hline
5x5 & 15  & 2 & 2 & 1.18 & - \\ \hline
15x15 & 5  & 6 & 1 & 1.00 & 1.07 \\ \hline
15x15 & 5  & 6 & 2 & 1.13 & 1.18 \\ \hline
\end{tabular}
\caption{$\chi^2/N$ against all the datasets 
used (H1, ZEUS, BCDMS) for variety of ENVPDF runs. }
\label{envpdftab}
\end{table}

%%%%%%%%%%%%%%%%%%%%%%%%%%%%%%%%%%%%%%%%%%%%%%%%%%%%%%%%%%%%%%%%%%%%%%%

\begin{figure}[h]
\begin{center}
\vspace{-0.2cm}
\hspace*{-1.5cm}
\epsfysize=8cm\epsffile{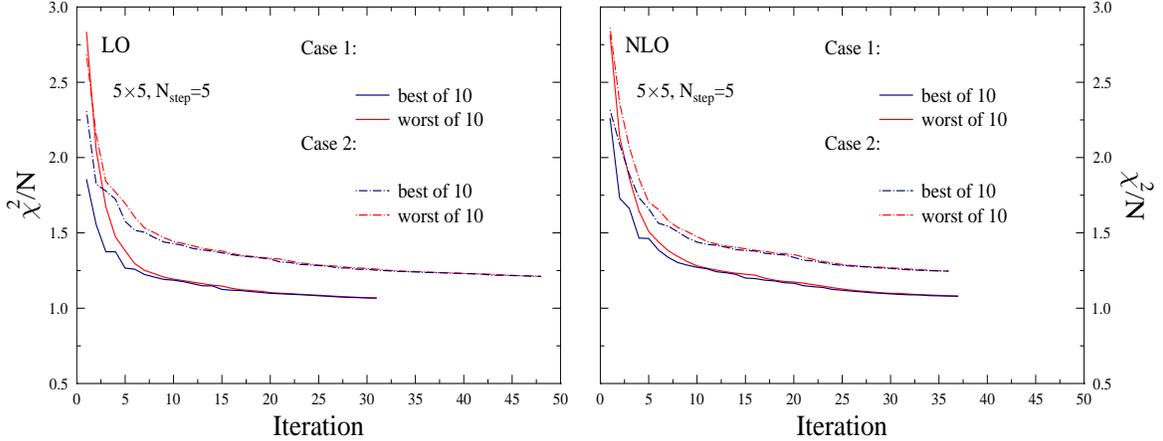} 
\vspace{-1.5cm}
\caption[a]{\protect \small (Colour online)
$\chi^2/N$ of the ENVPDF runs as a function
of the iteration.}
\label{chi2_ngenpdf.eps}
\end{center}
\end{figure}

Fig.~\ref{chi2_ngenpdf.eps} shows the $\chi^2/N$ as a 
function of iteration for 5x5 LO and NLO, both Case 1 and Case 2, runs, where
$N_{\rm step}=5$ and $N_{\rm orig}=2$. Clearly the ENVPDF runs take multiple 
number
of iterations for the $\chi^2/N$ to level compared to the MIXPDF runs, and 
they are therefore more costly in time. With the ENVPDF algorithm, however,   
the $\chi^2/N$ keeps on slowly
improving even after all the mother PDFs from the same iteration are equally 
good fits. For a larger 15x15 SOM the number of needed iterations remains
as large.

%%%%%%%%%%%%%%%%%%%%%%%%%%%%%%%%%%%%%%%%%%%%%%%%%%%%%%%%%%%%%%%%%%%%%%
\vspace{-0.0cm}
%%%%%%%%%%%%%%%%%%%%%%%%%%%%%%%%%%%%%%%%%%%%%%%%%%%%%%%%%%%%%%%%%%%%%%%

\begin{figure}[h]
\begin{center}
\vspace{-0.2cm}
\hspace*{-1.0cm}
\epsfysize=9cm\epsffile{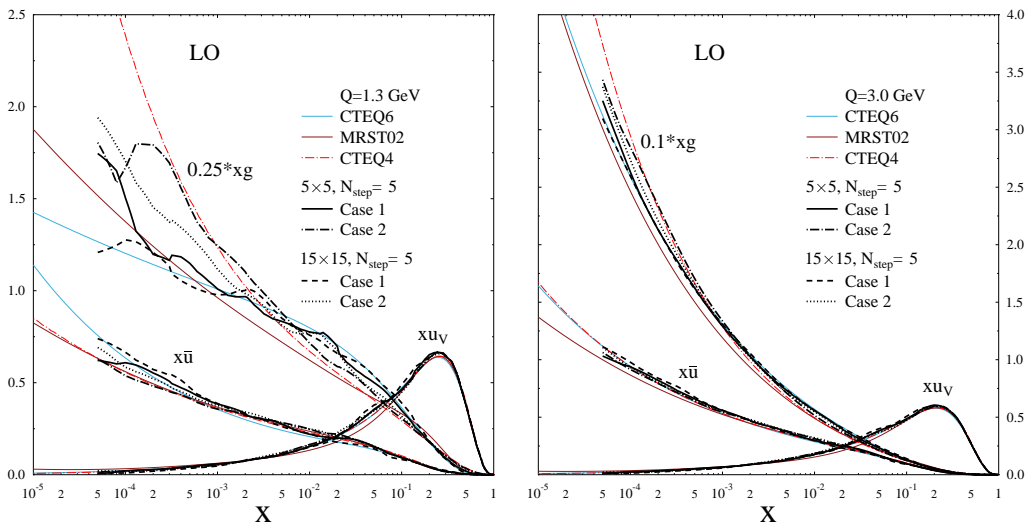} \hspace{-0.0cm}
\vspace{-1.5cm}
\caption[a]{\protect \small (Colour online)
LO ENVPDF results at the initial scale, and
at $Q=3.0$ GeV.}
\label{envpdf_jakaumat_lo.eps}
\end{center}
\end{figure}

\begin{figure}[h]
\begin{center}
\vspace{-0.2cm}
\hspace*{-1.0cm}
\epsfysize=9cm\epsffile{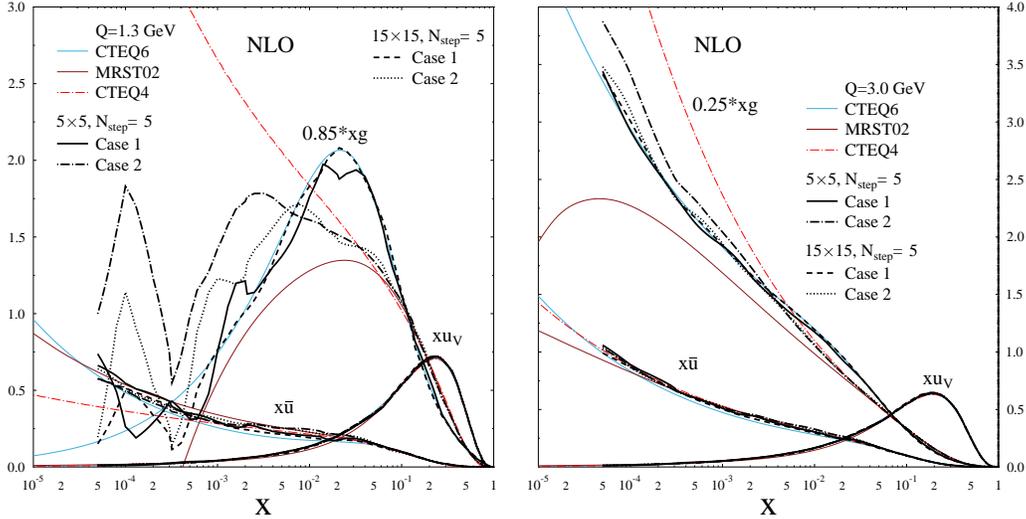} \hspace{-0.0cm}
\vspace{-1.5cm}
\caption[a]{\protect \small (Colour online)
NLO ENVPDF results at the initial scale, and
at $Q=3.0$ GeV.}
\label{envpdf_jakaumat_nlo.eps}
\end{center}
\end{figure}

Fig.~\ref{envpdf_jakaumat_lo.eps} shows some of the Case 1
LO ENVPDF results at the initial scale $Q=1.3$ GeV (left panel), and 
evolved up to $Q=3$ GeV (right panel). The reference curves shown are also 
evolved as in MIXPDF.
Although the initial scale ENVPDF 
results appear wiggly, they
smooth out soon because of the additional well known effect of QCD evolution. 
In fact, the initial scale curves could be 
made smoother
by applying a stronger Gaussian smoothing, but this is not necessary, 
as long as the starting scale is below the $Q^{\rm min}$ of the data. 
The evolved curves preserve the
initially set baryon number scaling within
$~0.5\%$ and momentum sumrule  within $~1.5\%$ accuracy.
Also, the results obtained
from a larger map tend to be smoother since the map PDFs get averaged with
a larger number of other PDFs. Studying the relation between the redundant
wiggliness of our initial scale PDFs and possible fitting of statistical 
fluctuations of the experimental data is beyond the scope of this paper.
The NLO Case 1 and 2 results are presented in 
Fig.~\ref{envpdf_jakaumat_nlo.eps}.
The trend of the results is clearly the same as in MIXPDF case, CTEQ6 is a 
favoured set, and especially the PDFs with gluons similar to those of CTEQ6's
have good $\chi^2/N$. 

We did not study the effect of modifying the width or the shape of the
envelope in detail here, but choosing the envelope to be the wider or narrower
than $1-\sigma$ for the Gaussian generate seem to lead both slower and poorer
convergence. Also, since we are
clustering on the similarity of the observables, the same cell may in theory
contain the best PDF of the iteration and PDFs which have $\chi^2/N$ as large
as 10. Therefore the shape of the envelope should be determined only by the 
curves with promising shapes.

%%%%%%%%%%%%%%%%%%%%%%%%%%%%%%%%%%%%%%%%%%%%%%%%%%%%%%%%%%%%%%%%%%%%%%%
\vspace{-0.0cm}
%%%%%%%%%%%%%%%%%%%%%%%%%%%%%%%%%%%%%%%%%%%%%%%%%%%%%%%%%%%%%%%%%%%%%%%

%%%%%%%%%%%%%%%%%%%%%%%%%%%%%%%%%%%%%%%%%%%%%%%%%%%%%%%%%%%%%%%%%%%%%%%
\vspace{-0.0cm}
%%%%%%%%%%%%%%%%%%%%%%%%%%%%%%%%%%%%%%%%%%%%%%%%%%%%%%%%%%%%%%%%%%%%%%%

\begin{figure}[h]
\begin{center}
\vspace{-0.2cm}
\hspace*{-1.0cm}
\epsfysize=9cm\epsffile{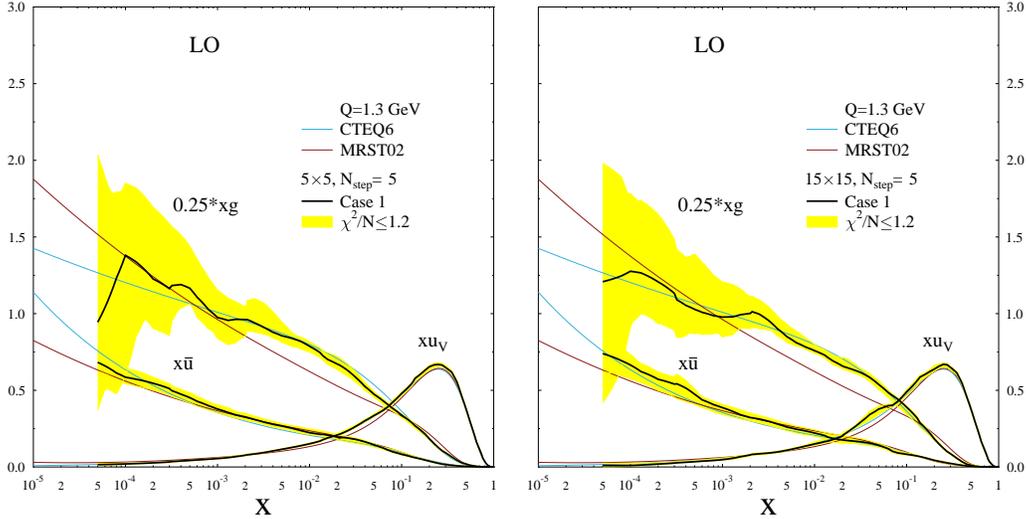} \hspace{-0.0cm}
\vspace{-1.5cm}
\caption[a]{\protect \small (Colour online)
LO ENVPDF best result and the  $\chi^2/N\le 1.2$
spread of results.  }
\label{envpdf_jakaumat_case1_good.eps}
\end{center}
\end{figure}

\begin{figure}[h]
\begin{center}
\vspace{-0.2cm}
\hspace*{-1.0cm}
\epsfysize=9cm\epsffile{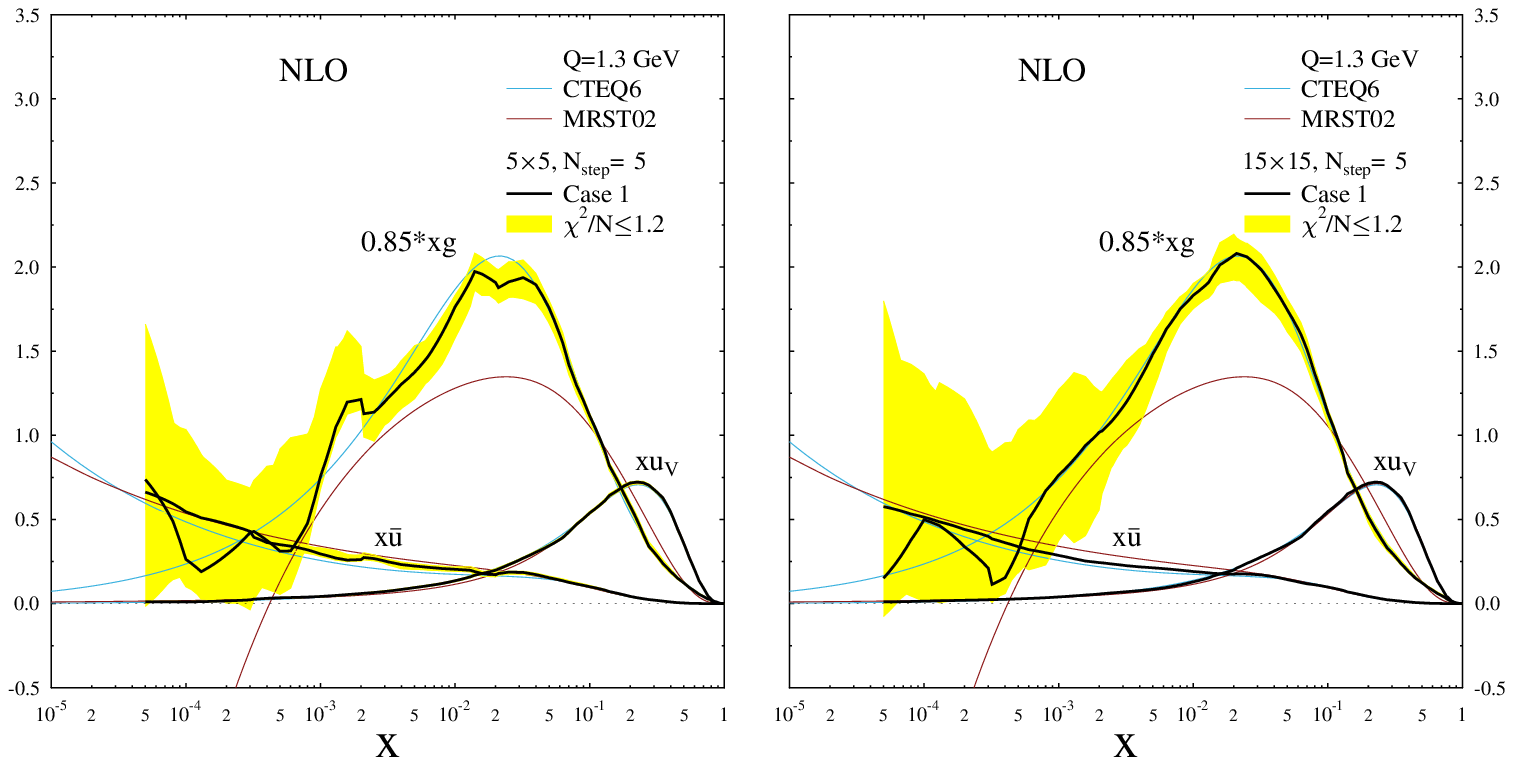} \hspace{-0.0cm}
\vspace{-1.5cm}
\caption[a]{\protect \small (Colour online)
NLO ENVPDF best result and the  $\chi^2/N\le 1.2$
spread of results. }
\label{envpdf_jakaumat_case1_good_nlo.eps}
\end{center}
\end{figure}

Next we want to study the PDF uncertainty using the unique
means the SOMs provide to us even for a case of PDFs without a functional form.
 Since we have only used DIS data in this 
introductory study, we are 
only able to explore the small-$x$ uncertainty for now.
Figs.~\ref{envpdf_jakaumat_case1_good.eps} (LO) and
\ref{envpdf_jakaumat_case1_good_nlo.eps} (NLO) showcase, besides our best 
results, the spread of all the initial scale
PDFs with $\chi^2/N\le 1.2$, that were
obtained during a $5\times 5$ (left panel)
and  $15\times 15$ (right panel) SOM run. Since the number of such PDF sets
is typically  of the order of thousands,
we only plot the minimum and maximum of 
the bundle of curves.
Since the total number of experimental 
datapoints used is $\sim$ 710, the spread 
$\Delta\chi^2/N\sim 0.2$ corresponds to
a $\Delta\chi^2\sim 140$. 
Expectedly, the small-$x$ gluons obtain the 
largest uncertainty for all the cases we studied.  Even though a
larger SOM with a larger database might be expected to have
 more variety in the shapes of the PDFs, 
the  $\chi^2/N\le 1.2$ spreads of the $5\times 5$ 
and  $15\times 15$ SOMs are 
more or less equal sized (the apparent differences in sizes
 at $Q=Q_0$ even out when the curves are evolved). 
Both maps therefore end up producing the same
extreme shapes for the map PDFs although a larger map has more subclasses
for them.
Remarkably then, a single SOM run can provide a quick uncertainty estimate
for a chosen  $\Delta\chi^2$ without performing a separate error analysis.

%%%%%%%%%%%%%%%%%%%%%%%%%%%%%%%%%%%%%%%%%%%%%%%%%%%%%%%%%%%%%%%%%%%%%%%
\vspace{-0.0cm}
%%%%%%%%%%%%%%%%%%%%%%%%%%%%%%%%%%%%%%%%%%%%%%%%%%%%%%%%%%%%%%%%%%%%%%%

%%%%%%%%%%%%%%%%%%%%%%%%%%%%%%%%%%%%%%%%%%%%%%%%%%%%%%%%%%%%%%%%%%%%%%%
\vspace{-0.0cm}
%%%%%%%%%%%%%%%%%%%%%%%%%%%%%%%%%%%%%%%%%%%%%%%%%%%%%%%%%%%%%%%%%%%%%%%

\begin{figure}[h]
\begin{center}
\vspace{-0.2cm}
\hspace*{-1.0cm}
\epsfysize=9cm\epsffile{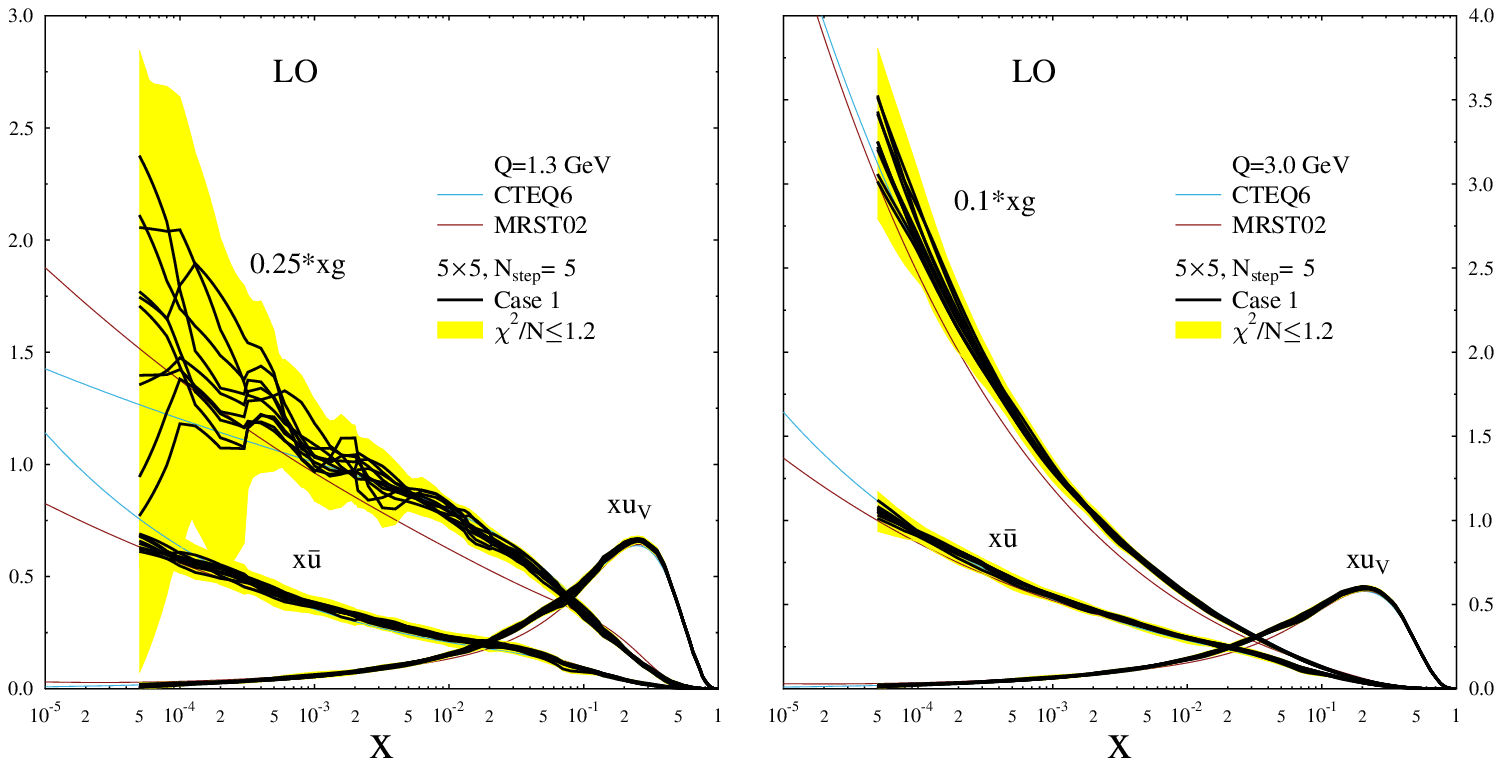} \hspace{-0.0cm}
\vspace{-1.5cm}
\caption[a]{\protect \small (Colour online)
 LO ENVPDF best results and the  $\chi^2/N\le 1.2$
spreads of results from 10 separate runs.}
\label{envpdf_jakaumat_case1_best.eps}
\end{center}
\end{figure}

\begin{figure}[h]
\begin{center}
\vspace{-0.2cm}
\hspace*{-1.0cm}
\epsfysize=9cm\epsffile{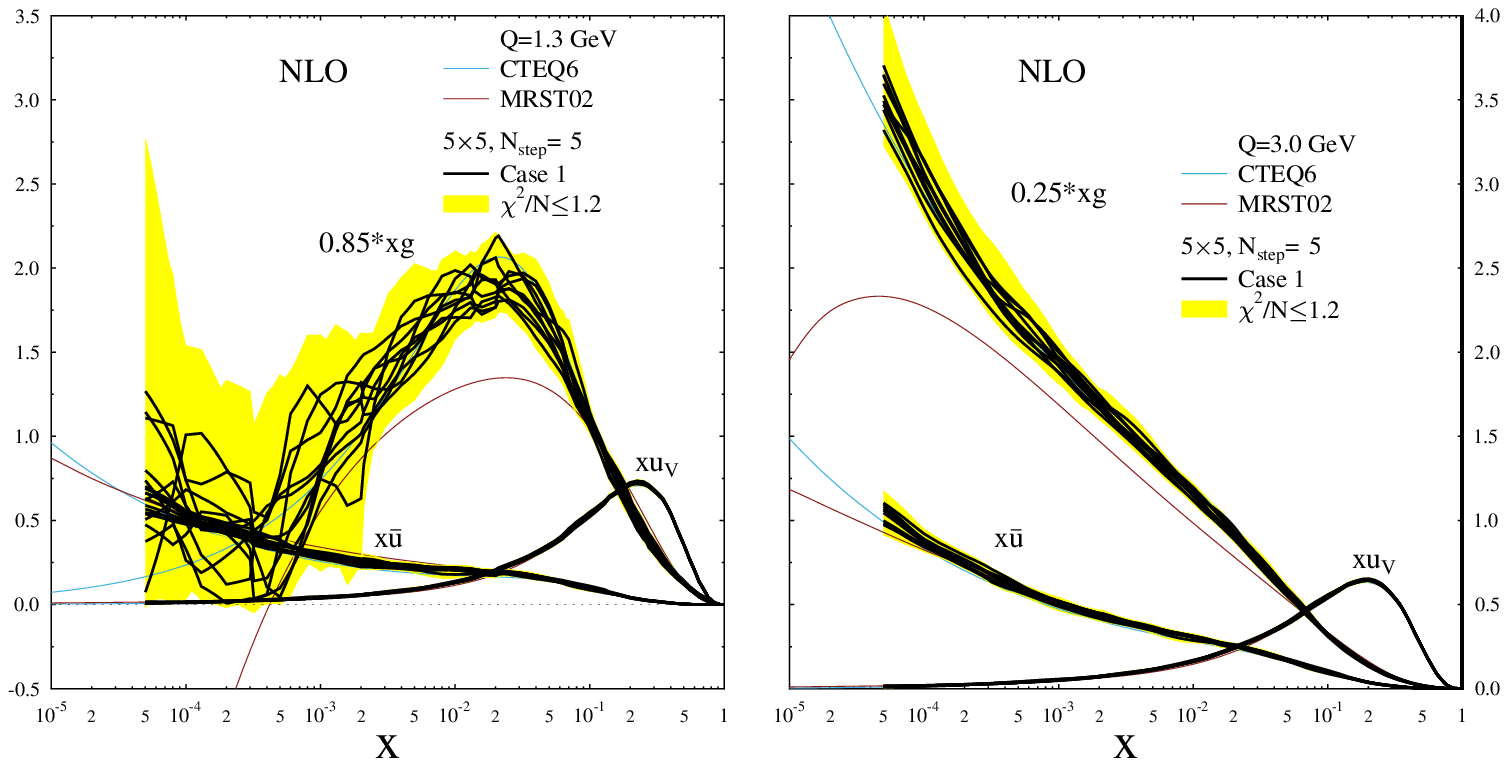} \hspace{-0.0cm}
\vspace{-1.5cm}
\caption[a]{\protect \small  (Colour online)
NLO ENVPDF best results and the  $\chi^2/N\le 1.2$
spreads of results from 10 separate runs.}
\label{envpdf_jakaumat_case1_best_nlo.eps}
\end{center}
\end{figure}

Due to the stochastical nature of the ENVPDF algorithm, we may
well also study the combined results from several separate runs.
It is especially important to verify the stability of our results, to show 
that the results are indeed reproducible instead of lucky coincidences.
Left panels of Figs~\ref{envpdf_jakaumat_case1_best.eps} (LO) and
\ref{envpdf_jakaumat_case1_best_nlo.eps} (NLO) present the best results, and 
the combined $\chi^2/N\le 1.2$ spreads for
10 repeated $5\times 5$, $N_{\rm step}=5$ runs at the initial scale. 
The average $\chi^2/N$ and the standard deviation $\sigma$ for these runs
are in LO (NLO) are 1.065 and 0.014 (1.122 and 0.029), corresponding
to $\Delta\chi^2\sim 10$ (20) for LO (NLO). 
The right panels of
the same  Figs~\ref{envpdf_jakaumat_case1_best.eps},
\ref{envpdf_jakaumat_case1_best_nlo.eps} show the 10 best result curves
and the  $\chi^2/N\le 1.2$ spreads evolved 
up to $Q=3.0$ GeV.
Clearly the seemingly large difference between the small-$x$ gluon results
at the initial scale is not statistically significant, but smooths out
when gluons are evolved.
Thus the initial
scale wiggliness of the PDFs is mainly only a residual effect from our method 
of generating  them and  not linked to the overtraining
of the SOM, and we refrain from studying cases where stronger initial
scale smoothing is applied.
Therefore our simple method of
producing the candidate PDFs by jittering random numbers inside a predetermined
envelope is surprisingly stable when used together with a complicated PDF 
processing that SOMs provide.

%%%%%%%%%%%%%%%%%%%%%%%%%%%%%%%%%%%%%%%%%%%%%%%%%%%%%%%%%%%%%%%%%%%%%%%
\vspace{-0.0cm}
%%%%%%%%%%%%%%%%%%%%%%%%%%%%%%%%%%%%%%%%%%%%%%%%%%%%%%%%%%%%%%%%%%%%%%%

%%%%%%%%%%%%%%%%%%%%%%%%%%%%%%%%%%%%%%%%%%%%%%%%%%%%%%%%%%%%%%%%%%%%%%%
\vspace{-0.0cm}
%%%%%%%%%%%%%%%%%%%%%%%%%%%%%%%%%%%%%%%%%%%%%%%%%%%%%%%%%%%%%%%%%%%%%%%

\section{Future of the SOMPDFs}
\label{future}

So far we have shown a relatively straightforward method of obtaining 
stochastically generated, parameter-free, PDFs, with an uncertainty estimate
for a desired $\Delta\chi^2$. On every iteration using our competitive learning
algorithm, 
the selection of the winning PDFs was based on the  $\chi^2/N$ alone, and
the fitting procedure was fully automated. In our MIXPDF algorithm the SOMs
were used merely as a tool to create new combinations, map PDFs, of our
input database. The ENVPDF algorithm also used the topology of the map to
determine the shape of the envelope, within which we sampled the database
PDFs. 

We reiterate that our initial study was aimed at observing and recording the behavior
of the SOM as an optimization tool. Many of the features of our
results could not in fact be predicted based on general assumptions.
The proposed method can be extended much further than that.
The automated version of the algorithm could be set to sample a vector
consisting of PDF parameters, instead of values of PDFs in each value of
$x$ of the data. That would lead to smooth, continuous type of solutions, 
either along the lines of global analyses, or NNPDFs using $N$ SOMs
for $N$ Monte-Carlo sampled replicas of the data.
Since the solution would be required to stay within an envelope of selected
width and shape given by the map, no restrictions for the parameters themselves
would be required. For such a method, all the existing error estimates, 
besides an uncertainty band produced by the map,
would be applicable as well.

What ultimately sets the SOM method apart from the standard global analyses or
NNPDF method, however, are the clustering and 
visualization possibilities that it offers. 
Instead of setting $M_{\mathrm{data}}=L_1$ and clustering according to the
similarity of the observables, it is possible to set the clustering criteria
to be anything that can be mathematically quantified, e.g. the shape of the
gluons or the large-$x$ behaviour of the PDFs. 
The desired feature of the PDFs can then be projected out from the SOM.
Moreover, by combining the method with an interactive graphic user 
interface (GUI), it
would be possible to change and control the shape and the width of the 
envelope as the
minimization proceeds, to guide the process by applying researcher insight at 
various stages of the process. Furthermore, 
the uncertainty band produced by the SOM as the 
run proceeds, could help the user to make decisions about the next steps of
the minimization.
With GUI it would be e.g. possible to constrain the extrapolation of the NN 
generated PDFs outside the $x$-range of the data without explicitly
introducing terms to ensure the correct small- and large-$x$ behaviour
as in NNPDF method (see Eq.(87) in \cite{Ball:2008by}).
The selection of the best PDF candidates for the subsequent iteration could 
then be made based on the user's preferences instead of solely based on the
 $\chi^2/N$.
That kind of method in turn could be extended
to multivariable cases such as nPDFs and even GPDs
and other not so well-known cases, where the data is too
sparse for stochastically generated, parameter-free, PDFs.

Generally, any PDF fitting method involves
a large number of
{\it flexible points} ``opportunities for adapting and fine tuning'', which
act as a source for both systematical and theoretical bias when fixed.
Obvious optimization
method independent sources of theoretical bias are the various 
parameters of the DGLAP equations, inclusion of extra sources of 
$Q^2$-dependence beyond DGLAP-type evolution and the data selection, affecting 
the coverage of different kinematical regions. SOMs themselves,
and different SOMPDF algorithm variations naturally also introduce flexible 
points of their own. We explored a little about the effects of choosing
the size of the SOM and the number of the batch training steps $N_{\rm step}$.
There are also plenty of other SOM properties that can be modified, such
as the shape of the SOM itself. We chose
to use a rectangular lattice, but generally the SOM can take any shape desired.
For demanding vizualisation purposes a hexagonal shape is an excellent choice,
since the meaning of the nearest neighbours is better defined.

The SOMPDF method, supplemented with the use of a GUI, will allow us to both  
qualitatively and quantitatively study the flexible points involved in the 
PDFs fitting. 
More complex hadronic matrix elements, such as the ones defining the GPDs, are 
natural candidates for future studies of cases where the experimental data 
are not numerous enough to allow
for a model independent fitting, and the guidance and intuition of the user is
therefore irreplaceable.
The method we are proposing is extremely open for user 
interaction, and the possibilities of such a method are widely unexplored.
\\\\

{\bf Acknowledgements}
\\ \\
We thank David Brogan for helping us to start this project.
This work was financially supported by the US
National Science Foundation grant no.0426971. 
HH was also supported by the U.S. Department of Energy, grant no.
DE-FG02-87ER40371. SL is supported by the U.S. Department of Energy, grant no.
DE-FG02-01ER41200.

\setlength{\baselineskip}{0.64cm}

\end{document}